\documentclass[12pt]{article}
\usepackage{verbatim}
\usepackage{amsfonts}
\usepackage{graphics}
\usepackage{amsmath}
\usepackage{times}
\usepackage{appendix}
\usepackage{graphicx}
\usepackage{color}
\usepackage{enumerate}
\usepackage{fancyhdr,latexsym,amsmath,amsfonts,amssymb,amsbsy,amsthm,url}
\usepackage[margin=0.5in,footskip=0.25in]{geometry}
\usepackage{graphics,graphicx,epsfig}
\usepackage{breqn}
\usepackage{subcaption}

\usepackage[english]{babel}
\usepackage[dvipsnames]{xcolor}
\usepackage{tikz}

\numberwithin{equation}{section}

\fancypagestyle{usual}{
	\lhead[\thepage]{}
	\chead[{\it \shortauthors}]{\sc \shorttitle}
	\rhead[]{\thepage}
	\lfoot[]{}
	\cfoot[]{}
	\rfoot[]{}
	
}

\makeatletter

\renewcommand{\section}{
	\@startsection
	{section}
	{1}
	{0pt}
	{1.1\baselineskip}
	{0.2\baselineskip}
	{\sc \centering}
}

\renewcommand{\subsection}{
	\@startsection
	{subsection}
	{1}
	{0pt}
	{1.1\baselineskip}
	{0.2\baselineskip}
	{\sc \centering}
}

\renewcommand{\subsubsection}{
	\@startsection
	{subsubsection}
	{1}
	{0pt}
	{1.1\baselineskip}
	{0.2\baselineskip}
	{\sc \centering}
}

\makeatother

\begin{document}
	
\title{\large\sc Modeling the dynamics of COVID-19 transmission in India: Social Distancing, Regional Spread and Healthcare Capacity}
\normalsize
\author{\sc{Suryadeepto Nag} \thanks{Indian Institute of Science Education and Research Pune, Pune-411008, Maharashtra, India, e-mail: suryadeepto.nag@students.iiserpune.ac.in}
\and \sc{Siddhartha P. Chakrabarty} \thanks{Department of Mathematics, Indian Institute of Technology Guwahati, Guwahati-781039, Assam, India, e-mail: pratim@iitg.ac.in}}
	
\date{\today}
\maketitle
\begin{abstract}
In the new paradigm of health-centric governance, policy makers are in a constant need for appropriate metrics and estimates in order to determine the best policies in a non-arbitrary fashion. Thus, in this paper, a compartmentalized model for the transmission of COVID-19 is developed to facilitate policy making. A socially distanced compartment is added to the model and its utility in quantifying the magnitude of voluntary social distancing is illustrated. Modifications are made to incorporate inter-region migration, and suitable metrics are proposed to quantify the impact of migration on the rise of cases. The healthcare capacity is modeled and a method is developed to study the consequences of the saturation of the healthcare system. The model and related measures are used to study the nature of the transmission and spread of COVID-19 in India, and appropriate insights are drawn.
\end{abstract}

	
\section{Introduction}
\label{Section_Introduction}

There are not many disease outbreaks in the collective memory of mankind that have had a global impact, to the extent that COVID-19 has. Therefore, the consequent attention of policy makers, towards mitigating the pandemic is hardly surprising to say the least. Even India, which as a country that has battled several disease outbreaks like malaria \cite{Sharma2015}, tuberculosis \cite{Pai2017} and water borne diseases \cite{Lakshminarayanan2015}, historically (and till date), a significant attention has naturally been accorded towards restricting the spread of COVID-19 in India. The various policies involved included periodic lockdowns, restrictions on business hours and public gatherings, and the closure of educational institutions \cite{Sharma2020}. India's first nationwide lockdown came on 24th March, the day on which the total number of cases in India crossed $500$ \cite{DataIndia}. Given the numbers, and put in context with India's population and population density, this precautionary move came early, when compared to other countries in Europe or North America, which had many more cases per capita, at that time \cite{Roser2020}, as compared to India. While the economic consequences of the lockdown in India were significant, \cite{Gupta2021A, Gupta2021B, Kapoor2021}, it is undeniable, that lockdowns, by their design, reduce the rate of spread of the infection by reducing the frequency of contact between individuals \cite{Di2020}, thereby reducing the number of deaths per unit time. While it may not necessarily reduce the total number of deaths in a population \cite{Gibson2020}, it may help in delaying the peak, without altering the area under the graph. Lockdowns have other benefits, as it provides sufficient time for the healthcare system to be better organized and equipped to deal with the pandemic. Lowering the rate of transmission may also be particularly beneficial when the number of cases requiring hospitalization is nearing the healthcare capacity. Thus, it is crucial to develop a systematic method to determine what is an appropriate time to impose lockdowns. Such a time will depend on other quantifiable measures of the extent of social distancing in the absence of lockdowns, caseload of the healthcare system and the overall resilience of the system. Unlike physical distancing, which involves people maintaining a certain distance from people around them, social distancing involves people avoiding going out or meeting and gathering with people unless essential. Voluntary social distancing can reduce the rate of transmission of the infection by reducing interactions between individuals, and thus if the rate of voluntary social distancing is high, a lockdown may not be necessary. Motivated by these factors, we develop two specific methods to estimate suitable parameters that can guide policymakers in imposing lockdowns or other civil restrictions, the first being the existing rate of social distancing and the second being the maximum healthcare capacity, along with the consequences of exceeding the said capacity.

Additionally, apart from lockdowns and civil restrictions within regions, another method used by governments of the world, is travel restrictions. This has been repeatedly practiced at an international level to curb the cases of new infections spread from inbound infectious individuals. During the lockdowns in India, there were often inter-state and inter-district travel restrictions. These restrictions can cause losses of economic opportunities for individuals, communities and firms. It may also lead to logistical issues for individuals who may need to travel for the sake of emergencies or subsistence. Thus, it is of paramount interest for policy makers to have a measure of the impact of allowing/restricting inter-region migration on the number of cases in the region.

In this paper, we develop a modified version of the standard SEIR model, which conventionally includes compartments of susceptible, exposed, infectious and removed individuals. We modify it to include a class of individuals who are voluntarily social distancing, in order to be able to quantify the number of people moving in and out of social distancing on a given date. We also subsequently make modifications to the model to account for regional migration, thus treating the system as a network of connected systems rather than a well-mixed homogeneous one. This allows us to develop appropriate quantifiers that may guide governments on taking appropriate policy decisions surrounding inter-region migration. Lastly, we modify our model suitably to factor in the maximum hospital capacity and its effect on the overall death rate of the system.
	
\section{Materials and Methods}
\label{Section_Methods}

\subsection{Data}
\label{Subsection_Data}
	
We use data from primarily three sources. For COVID-19 data, we use data from the website of COVID-19 tracker in India \cite{DataIndia}. It is a publicly available dataset on the spread of the COVID-19 pandemic in India. It also has data on daily new cases, recoveries and deaths, for 36 states and union territories in India. We use the statewise data, as well as the aggregate numbers for the countries. Apart from data on infections, we also use data on populations of states and the country \cite{PopIndia}. This dataset has the statewise population projections for the year 2020. The population data is used to find the value of the population parameter $N$ defined in subsequent sections. It is also used to find the per-capita cases for the different states in order to study the statewise spread. We also use tourism data in India from statistics published by the Ministry of Tourism, Government of India \cite{TIS2020, TIS2021}. ``India Tourism Statistics at a Glance 2021'' has data on the number of tourists visiting the 10 states most frequented by tourists. We can compare these numbers with those from ``India Tourism Statistics 2020" which has the numbers for 2019. We can then use these statistics to study the effect of travel and tourism on the spread of COVID-19 to different regions.
	
\subsection{Model}

In this part we describe the models that we have considered for our study. To begin with we consider a compartment based model which takes into account the act of voluntary social distancing as well as isolation. The second model that we consider is in the paradigm of travel restrictions, particularly to address questions pertaining to the timing of imposition of such restrictions. Finally, we consider a model setup for the capacity of the healthcare system in dealing with the pandemic driven health support requirements.
	
\subsubsection{Compartmentalized Model Incorporating Voluntary Social Distancing}
\label{Subsection_SEIRQ_model}
	
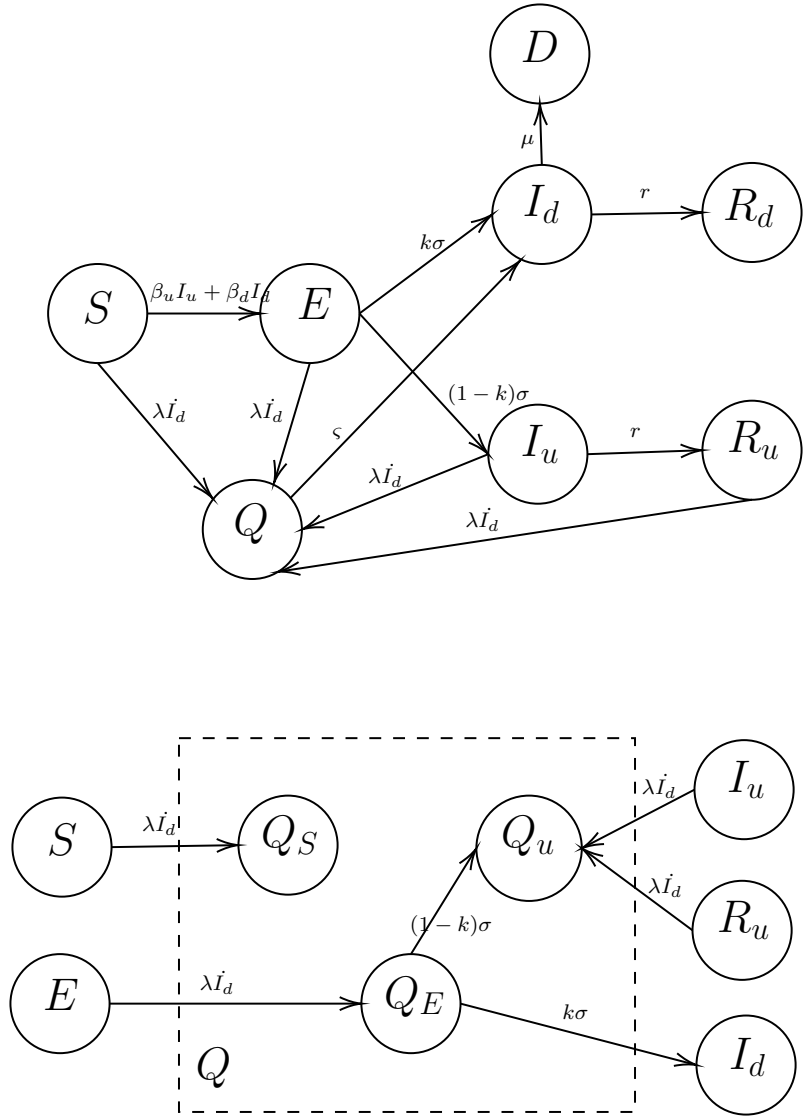
\begin{figure}[h]
\begin{center}
\tikzset{every picture/.style={line width=0.75pt}} 
\tikzset{every picture/.style={line width=0.75pt}} 
\tikzset{every picture/.style={line width=0.75pt}} 
\tikzset{every picture/.style={line width=0.75pt}} 
\begin{tikzpicture}[x=0.75pt,y=0.75pt,yscale=-1,xscale=1]
				\draw   (139,166) .. controls (139,152.19) and (150.19,141) .. (164,141) .. controls (177.81,141) and (189,152.19) .. (189,166) .. controls (189,179.81) and (177.81,191) .. (164,191) .. controls (150.19,191) and (139,179.81) .. (139,166) -- cycle ;
				\draw   (246,166) .. controls (246,152.19) and (257.19,141) .. (271,141) .. controls (284.81,141) and (296,152.19) .. (296,166) .. controls (296,179.81) and (284.81,191) .. (271,191) .. controls (257.19,191) and (246,179.81) .. (246,166) -- cycle ;
				\draw   (362,35) .. controls (362,21.19) and (373.19,10) .. (387,10) .. controls (400.81,10) and (412,21.19) .. (412,35) .. controls (412,48.81) and (400.81,60) .. (387,60) .. controls (373.19,60) and (362,48.81) .. (362,35) -- cycle ;
				\draw   (361,237) .. controls (361,223.19) and (372.19,212) .. (386,212) .. controls (399.81,212) and (411,223.19) .. (411,237) .. controls (411,250.81) and (399.81,262) .. (386,262) .. controls (372.19,262) and (361,250.81) .. (361,237) -- cycle ;
				\draw   (363,116) .. controls (363,102.19) and (374.19,91) .. (388,91) .. controls (401.81,91) and (413,102.19) .. (413,116) .. controls (413,129.81) and (401.81,141) .. (388,141) .. controls (374.19,141) and (363,129.81) .. (363,116) -- cycle ;
				\draw   (469,115) .. controls (469,101.19) and (480.19,90) .. (494,90) .. controls (507.81,90) and (519,101.19) .. (519,115) .. controls (519,128.81) and (507.81,140) .. (494,140) .. controls (480.19,140) and (469,128.81) .. (469,115) -- cycle ;
				\draw   (469,235) .. controls (469,221.19) and (480.19,210) .. (494,210) .. controls (507.81,210) and (519,221.19) .. (519,235) .. controls (519,248.81) and (507.81,260) .. (494,260) .. controls (480.19,260) and (469,248.81) .. (469,235) -- cycle ;
				\draw   (217,275) .. controls (217,261.19) and (228.19,250) .. (242,250) .. controls (255.81,250) and (267,261.19) .. (267,275) .. controls (267,288.81) and (255.81,300) .. (242,300) .. controls (228.19,300) and (217,288.81) .. (217,275) -- cycle ;
				\draw    (189,166) -- (244,166) ;
				\draw [shift={(246,166)}, rotate = 180] [color={rgb, 255:red, 0; green, 0; blue, 0 }  ][line width=0.75]    (10.93,-3.29) .. controls (6.95,-1.4) and (3.31,-0.3) .. (0,0) .. controls (3.31,0.3) and (6.95,1.4) .. (10.93,3.29)   ;
				\draw    (164,191) -- (220.71,257.81) ;
				\draw [shift={(222,259.33)}, rotate = 229.68] [color={rgb, 255:red, 0; green, 0; blue, 0 }  ][line width=0.75]    (10.93,-3.29) .. controls (6.95,-1.4) and (3.31,-0.3) .. (0,0) .. controls (3.31,0.3) and (6.95,1.4) .. (10.93,3.29)   ;
				\draw    (271,191) -- (253.57,249.42) ;
				\draw [shift={(253,251.33)}, rotate = 286.61] [color={rgb, 255:red, 0; green, 0; blue, 0 }  ][line width=0.75]    (10.93,-3.29) .. controls (6.95,-1.4) and (3.31,-0.3) .. (0,0) .. controls (3.31,0.3) and (6.95,1.4) .. (10.93,3.29)   ;
				\draw    (296,166) -- (361.4,117.2) ;
				\draw [shift={(363,116)}, rotate = 503.27] [color={rgb, 255:red, 0; green, 0; blue, 0 }  ][line width=0.75]    (10.93,-3.29) .. controls (6.95,-1.4) and (3.31,-0.3) .. (0,0) .. controls (3.31,0.3) and (6.95,1.4) .. (10.93,3.29)   ;
				\draw    (296,166) -- (359.65,235.52) ;
				\draw [shift={(361,237)}, rotate = 227.53] [color={rgb, 255:red, 0; green, 0; blue, 0 }  ][line width=0.75]    (10.93,-3.29) .. controls (6.95,-1.4) and (3.31,-0.3) .. (0,0) .. controls (3.31,0.3) and (6.95,1.4) .. (10.93,3.29)   ;
				\draw    (413,116) -- (467,115.04) ;
				\draw [shift={(469,115)}, rotate = 538.98] [color={rgb, 255:red, 0; green, 0; blue, 0 }  ][line width=0.75]    (10.93,-3.29) .. controls (6.95,-1.4) and (3.31,-0.3) .. (0,0) .. controls (3.31,0.3) and (6.95,1.4) .. (10.93,3.29)   ;
				\draw    (388,91) -- (387.06,62) ;
				\draw [shift={(387,60)}, rotate = 448.15] [color={rgb, 255:red, 0; green, 0; blue, 0 }  ][line width=0.75]    (10.93,-3.29) .. controls (6.95,-1.4) and (3.31,-0.3) .. (0,0) .. controls (3.31,0.3) and (6.95,1.4) .. (10.93,3.29)   ;
				\draw    (411,237) -- (467,235.07) ;
				\draw [shift={(469,235)}, rotate = 538.03] [color={rgb, 255:red, 0; green, 0; blue, 0 }  ][line width=0.75]    (10.93,-3.29) .. controls (6.95,-1.4) and (3.31,-0.3) .. (0,0) .. controls (3.31,0.3) and (6.95,1.4) .. (10.93,3.29)   ;
				\draw    (361,237) -- (268.85,274.25) ;
				\draw [shift={(267,275)}, rotate = 337.99] [color={rgb, 255:red, 0; green, 0; blue, 0 }  ][line width=0.75]    (10.93,-3.29) .. controls (6.95,-1.4) and (3.31,-0.3) .. (0,0) .. controls (3.31,0.3) and (6.95,1.4) .. (10.93,3.29)   ;
				\draw    (494,260) -- (256.98,296.03) ;
				\draw [shift={(255,296.33)}, rotate = 351.36] [color={rgb, 255:red, 0; green, 0; blue, 0 }  ][line width=0.75]    (10.93,-3.29) .. controls (6.95,-1.4) and (3.31,-0.3) .. (0,0) .. controls (3.31,0.3) and (6.95,1.4) .. (10.93,3.29)   ;
				\draw    (261,259.33) -- (375.61,140.77) ;
				\draw [shift={(377,139.33)}, rotate = 494.03] [color={rgb, 255:red, 0; green, 0; blue, 0 }  ][line width=0.75]    (10.93,-3.29) .. controls (6.95,-1.4) and (3.31,-0.3) .. (0,0) .. controls (3.31,0.3) and (6.95,1.4) .. (10.93,3.29)   ;
				\draw  [dash pattern={on 4.5pt off 4.5pt}] (205,380.33) -- (435,380.33) -- (435,569) -- (205,569) -- cycle ;
				\draw   (235,434.33) .. controls (235,420.53) and (246.19,409.33) .. (260,409.33) .. controls (273.81,409.33) and (285,420.53) .. (285,434.33) .. controls (285,448.14) and (273.81,459.33) .. (260,459.33) .. controls (246.19,459.33) and (235,448.14) .. (235,434.33) -- cycle ;
				\draw   (355,435.92) .. controls (355,421.24) and (366.9,409.33) .. (381.58,409.33) .. controls (396.26,409.33) and (408.17,421.24) .. (408.17,435.92) .. controls (408.17,450.6) and (396.26,462.5) .. (381.58,462.5) .. controls (366.9,462.5) and (355,450.6) .. (355,435.92) -- cycle ;
				\draw   (297,514.33) .. controls (297,500.53) and (308.19,489.33) .. (322,489.33) .. controls (335.81,489.33) and (347,500.53) .. (347,514.33) .. controls (347,528.14) and (335.81,539.33) .. (322,539.33) .. controls (308.19,539.33) and (297,528.14) .. (297,514.33) -- cycle ;
				\draw   (121,435.33) .. controls (121,421.53) and (132.19,410.33) .. (146,410.33) .. controls (159.81,410.33) and (171,421.53) .. (171,435.33) .. controls (171,449.14) and (159.81,460.33) .. (146,460.33) .. controls (132.19,460.33) and (121,449.14) .. (121,435.33) -- cycle ;
				\draw   (120,514.33) .. controls (120,500.53) and (131.19,489.33) .. (145,489.33) .. controls (158.81,489.33) and (170,500.53) .. (170,514.33) .. controls (170,528.14) and (158.81,539.33) .. (145,539.33) .. controls (131.19,539.33) and (120,528.14) .. (120,514.33) -- cycle ;
				\draw   (466,545.33) .. controls (466,531.53) and (477.19,520.33) .. (491,520.33) .. controls (504.81,520.33) and (516,531.53) .. (516,545.33) .. controls (516,559.14) and (504.81,570.33) .. (491,570.33) .. controls (477.19,570.33) and (466,559.14) .. (466,545.33) -- cycle ;
				\draw   (465,406.33) .. controls (465,392.53) and (476.19,381.33) .. (490,381.33) .. controls (503.81,381.33) and (515,392.53) .. (515,406.33) .. controls (515,420.14) and (503.81,431.33) .. (490,431.33) .. controls (476.19,431.33) and (465,420.14) .. (465,406.33) -- cycle ;
				\draw   (464,477.33) .. controls (464,463.53) and (475.19,452.33) .. (489,452.33) .. controls (502.81,452.33) and (514,463.53) .. (514,477.33) .. controls (514,491.14) and (502.81,502.33) .. (489,502.33) .. controls (475.19,502.33) and (464,491.14) .. (464,477.33) -- cycle ;
				\draw    (171,435.33) -- (233,434.36) ;
				\draw [shift={(235,434.33)}, rotate = 539.1] [color={rgb, 255:red, 0; green, 0; blue, 0 }  ][line width=0.75]    (10.93,-3.29) .. controls (6.95,-1.4) and (3.31,-0.3) .. (0,0) .. controls (3.31,0.3) and (6.95,1.4) .. (10.93,3.29)   ;
				\draw    (170,514.33) -- (295,514.33) ;
				\draw [shift={(297,514.33)}, rotate = 180] [color={rgb, 255:red, 0; green, 0; blue, 0 }  ][line width=0.75]    (10.93,-3.29) .. controls (6.95,-1.4) and (3.31,-0.3) .. (0,0) .. controls (3.31,0.3) and (6.95,1.4) .. (10.93,3.29)   ;
				\draw    (465,406.33) -- (409.94,434.99) ;
				\draw [shift={(408.17,435.92)}, rotate = 332.5] [color={rgb, 255:red, 0; green, 0; blue, 0 }  ][line width=0.75]    (10.93,-3.29) .. controls (6.95,-1.4) and (3.31,-0.3) .. (0,0) .. controls (3.31,0.3) and (6.95,1.4) .. (10.93,3.29)   ;
				\draw    (464,477.33) -- (409.77,437.11) ;
				\draw [shift={(408.17,435.92)}, rotate = 396.57] [color={rgb, 255:red, 0; green, 0; blue, 0 }  ][line width=0.75]    (10.93,-3.29) .. controls (6.95,-1.4) and (3.31,-0.3) .. (0,0) .. controls (3.31,0.3) and (6.95,1.4) .. (10.93,3.29)   ;
				\draw    (347,514.33) -- (464.06,544.83) ;
				\draw [shift={(466,545.33)}, rotate = 194.6] [color={rgb, 255:red, 0; green, 0; blue, 0 }  ][line width=0.75]    (10.93,-3.29) .. controls (6.95,-1.4) and (3.31,-0.3) .. (0,0) .. controls (3.31,0.3) and (6.95,1.4) .. (10.93,3.29)   ;
				\draw    (322,489.33) -- (353.95,437.62) ;
				\draw [shift={(355,435.92)}, rotate = 481.71] [color={rgb, 255:red, 0; green, 0; blue, 0 }  ][line width=0.75]    (10.93,-3.29) .. controls (6.95,-1.4) and (3.31,-0.3) .. (0,0) .. controls (3.31,0.3) and (6.95,1.4) .. (10.93,3.29)   ;
				\draw (155,153.73) node [anchor=north west][inner sep=0.75pt]  [font=\Large]  {$S$};
				\draw (263,151.73) node [anchor=north west][inner sep=0.75pt]  [font=\Large]  {$E$};
				\draw (231,260.73) node [anchor=north west][inner sep=0.75pt]  [font=\Large]  {$Q$};
				\draw (377,99.73) node [anchor=north west][inner sep=0.75pt]  [font=\Large]  {$I_{d}$};
				\draw (376,220.73) node [anchor=north west][inner sep=0.75pt]  [font=\Large]  {$I_{u}$};
				\draw (480,218.73) node [anchor=north west][inner sep=0.75pt]  [font=\Large]  {$R_{u}$};
				\draw (478,101.73) node [anchor=north west][inner sep=0.75pt]  [font=\Large]  {$R_d$};
				\draw (376,21.73) node [anchor=north west][inner sep=0.75pt]  [font=\Large]  {$D$};
				\draw (189,149.4) node [anchor=north west][inner sep=0.75pt]  [font=\scriptsize]  {$\beta _{u} I_{u} +\beta _{d} I_{d}$};
				\draw (325,124.4) node [anchor=north west][inner sep=0.75pt]  [font=\scriptsize]  {$k\sigma $};
				\draw (339,200.4) node [anchor=north west][inner sep=0.75pt]  [font=\scriptsize]  {$( 1-k) \sigma $};
				\draw (376,74.4) node [anchor=north west][inner sep=0.75pt]  [font=\scriptsize]  {$\mu $};
				\draw (190,207.4) node [anchor=north west][inner sep=0.75pt]  [font=\scriptsize]  {$\lambda \dot{I_{d}}$};
				\draw (239,207.4) node [anchor=north west][inner sep=0.75pt]  [font=\scriptsize]  {$\lambda \dot{I_{d}}$};
				\draw (299,240.4) node [anchor=north west][inner sep=0.75pt]  [font=\scriptsize]  {$\lambda \dot{I_{d}}$};
				\draw (348,262.4) node [anchor=north west][inner sep=0.75pt]  [font=\scriptsize]  {$\lambda \dot{I_{d}}$};
				\draw (431,222.4) node [anchor=north west][inner sep=0.75pt]  [font=\scriptsize]  {$r$};
				\draw (436,101.4) node [anchor=north west][inner sep=0.75pt]  [font=\scriptsize]  {$r$};
				\draw (281,222.4) node [anchor=north west][inner sep=0.75pt]  [font=\scriptsize]  {$\varsigma $};
				\draw (137,422.07) node [anchor=north west][inner sep=0.75pt]  [font=\Large]  {$S$};
				\draw (135,500.07) node [anchor=north west][inner sep=0.75pt]  [font=\Large]  {$E$};
				\draw (481,530.07) node [anchor=north west][inner sep=0.75pt]  [font=\Large]  {$I_{d}$};
				\draw (480,390.07) node [anchor=north west][inner sep=0.75pt]  [font=\Large]  {$I_{u}$};
				\draw (475,461.07) node [anchor=north west][inner sep=0.75pt]  [font=\Large]  {$R_{u}$};
				\draw (245,418.07) node [anchor=north west][inner sep=0.75pt]  [font=\Large]  {$Q_{S}$};
				\draw (307,499.07) node [anchor=north west][inner sep=0.75pt]  [font=\Large]  {$Q_{E}$};
				\draw (366,418.07) node [anchor=north west][inner sep=0.75pt]  [font=\Large]  {$Q_{u}$};
				\draw (212,535.07) node [anchor=north west][inner sep=0.75pt]  [font=\Large]  {$Q$};
				\draw (185,416.07) node [anchor=north west][inner sep=0.75pt]  [font=\scriptsize]  {$\lambda \dot{I_{d}}$};
				\draw (214,496.07) node [anchor=north west][inner sep=0.75pt]  [font=\scriptsize]  {$\lambda \dot{I_{d}}$};
				\draw (437,397.07) node [anchor=north west][inner sep=0.75pt]  [font=\scriptsize]  {$\lambda \dot{I_{d}}$};
				\draw (440,447.07) node [anchor=north west][inner sep=0.75pt]  [font=\scriptsize]  {$\lambda \dot{I_{d}}$};
				\draw (397,513.07) node [anchor=north west][inner sep=0.75pt]  [font=\scriptsize]  {$k\sigma $};
				\draw (320,468.07) node [anchor=north west][inner sep=0.75pt]  [font=\scriptsize]  {$( 1-k) \sigma $};
\end{tikzpicture}
\caption{Schematic representation of the mathematical model.}
\label{One_Schematic_Diagram}
\end{center}
\end{figure}

While several compartmentalized model which take in to account quarantined or isolated individuals, exist in literature \cite{Feng2007, Ali2020, Castilho2020, He2020, Gupta2021C}, there are hardly any models that factor in a compartment of individuals who are voluntarily social distancing. In order to model the spread of coronavirus and coronavirus-like epidemics, we begin with a simplified model of a closed system, with equal birth and death rate (this death rate is the number of deaths per capita of the system, excluding those caused due to the mortality effects of the disease in consideration). Our model also assumes that reinfections do not happen \textit{i.e.,} it is not possible for an individual who was infected, to get reinfected. While this may restrict the scope of our analysis, such that we may only consider periods when a single variant of the virus is active or periods suitably small such that the immunity of individuals are strong enough to prevent reinfections, this key assumption appeals to parsimony and prevents the model from becoming excessively complicated. Accordingly, as already noted, we start with a simplified model of a closed system with constant parameters. Both the assumptions of a closed system and constant parameters are partially relaxed in subsequent sections. The model takes into consideration the classical modeling approach of epidemics by incorporating the susceptible, exposed, infectious and removed individuals. This is then extended to incorporate the impact of social distancing and isolation. In addition, the infectious and the recovered compartments are each subdivided into two compartments, distinguishing between detected and undetected infections. Also the socially distanced compartment is divided into three compartments, with the final compartment being the deceased compartment. The schematic representation of the model as described is given in Figure \ref{One_Schematic_Diagram}.
We assume that the infectious population are of two types, namely, that which has been detected, and that which has not been detected. Furthermore, it has been assumed that the undetected cases are either mild or asymptomatic in nature, since it is reasonable to expect that cases which require medical attention will very likely be detected, due to visible symptoms. This simplification of assuming that the undetected cases are mild, is based on the premise that serious cases would have required and received medical treatment, and thus been detected. Therefore, in this model, undetected cases do not lead to deaths.

A major modification in our model, when compared to other existing models, is the presence of a socially distanced class. The socially distanced class is the subset of the population who are voluntarily social distancing. We assume that once individuals move into this class, they can no longer be exposed to the virus. Therefore, it naturally includes all the people who are (due to other reasons, like remote locations) not susceptible to exposure to begin with. However, individuals who have already been exposed, individuals who are infected but undiagnosed, and individuals who have recovered but their infection had never been detected, may still move into the socially distanced class. Only individuals with detected infections, and those recovered from detected infections, do not move into the social distancing class. This is because, we define social distancing as a preemptive voluntary measure that individuals practice to prevent themselves from contracting the illness. The individuals whose infections have been detected, do not distance because they cannot be reinfected again. It is possible that individuals who have tested positive for the illness will be responsible and quarantine. But their dynamics are different from those individuals distancing preemptively, and we assume that the ``infectious detected'' class represents the average behavior of individuals whose infections have been detected (as accounting for behavior patterns will introduce more compartments into what is already a complex system). The dynamics of voluntary social-distancing to avoid infections is explained below in the context of the model.

As seen from Figure \ref{One_Schematic_Diagram}, the modeling representation can be made in two parts. Although this two part setup is equivalent to a single system, wherein each compartment interacts independently, it is useful to classify $Q_{S}$ (socially distanced susceptible), $Q_{E}$ (socially distanced exposed) and $Q_{u}$ (socially distanced unidentified), under the umbrella of the ``socially distanced class'' $Q$ (We stick with $Q$ to describe the variable, following literature on SEIRQ models). 
\begin{enumerate}[(i)]
\item In the absence of social distancing, a susceptible individual ($S$) becomes exposed ($E$). Once exposed, after the incubation period, the individual may either be asymptomatic or develop a mild illness, and their infection may remain undetected ($I_{u}$) or they may get tested positive for the infection ($I_{d}$). If they recover, they, respectively, move to the classes of undetected recovered ($R_{u}$) or detected recovered ($R_{d}$). For some cases of serious infections in $I_d$, the case may lead to a fatality and the individual goes to the death ($D$) compartment. 
\item However, due to widespread public apprehension, resulting primarily from progressively rapid rise in case loads, as well as restrictions imposed by the health authorities, individuals may adopt the practice of social distancing. The susceptible individuals go on to the $Q_{S}$ compartment, and are safe as long as they continue to be in $Q_{S}$. Since exposed individuals are not aware that they have been already exposed to the virus, they may apprehensively move to the $Q_{E}$ compartment. In case they develop symptoms and test positive, they move to $I_{d}$, otherwise they will join other undiagnosed infected persons who (unaware of their infection) preemptively practice social distancing in $Q_{u}$. While individuals in $Q_{u}$ may be infected, they cannot spread the infection as they are themselves in social distancing. Undiagnosed yet recovered individuals may also want to social distance as they are unaware of their (now, cured) infection, but since it does not make a difference to the model (they cannot infect), we can group them with $Q_{u}$. Note that in Figure \ref{One_Schematic_Diagram}, the outflow from $Q$ to $I_{d}$ shown as $\varsigma Q$, is actually the flow shown from $Q_{E}$ to $I_{d}$ which is $k\sigma Q_{E}$.
\end{enumerate}
In order to model the voluntary social distancing, we assume that the rate with which people move in and out of social distancing is proportional to the rate of change of active cases. If there is a surge of new cases, then people are likely to be apprehensive and adopt greater precautions, while if there are few cases and a large number of recoveries, people are more likely to venture out rather than stay at home. This, however, applies to those people who have not contracted the disease yet (to their knowledge). If $X$ is a compartment and $Q_{X}$ is the corresponding compartment of those in $X$ who have socially distanced, then we model the flow by,
\begin{equation}
\dot{Q_{X}}=\lambda \dot{I_{d}} X,
\end{equation}
where $\lambda$ is a constant and $I_{d}$ is the number of detected active infections. Such a relationship implies that as the number of active diagnosed cases increases (positive $\dot{I_d}$), more individuals will isolate themselves in apprehension, and as the number of cases decrease, individuals will stop isolating themselves. Obviously, the rate with which people isolate will depend only on detected cases. This model works when the number of individuals in the socially distanced class $Q$ is large. $Q$ being large, stops the model from running into a scenario in which the number of people adopting social distancing is negative, since it has no meaning physically. Therefore, this model works in the situation where there are always a few people social distancing. As discussed earlier, the rate of people moving to the social distancing class depends both on government restrictions and on the public perception of the risk. In our formulation, government restrictions are implicitly incorporated in the magnitude of $\lambda$ and the perception of the risk is modeled to linearly depend on the number of new detected cases. The mathematical model for the schematic representation in Figure \ref{One_Schematic_Diagram}, is given by the following system of nonlinear ordinary differential equations (ODEs),
\begin{eqnarray}
\dot{S}&=&-(\beta_{u}I_{u}+\beta_{d}I_{d})S-\lambda\dot{I_{d}}S \nonumber \\
\dot{E}&=&(\beta_{u}I_{u}+\beta_{d}I_{d})S-\sigma E-\lambda\dot{I_{d}}E \nonumber \\
\dot{I_{u}}&=&(1-k)\sigma E-rI_{u}-\lambda\dot{I_{d}}I_{u} \nonumber \\
\dot{I_{d}}&=&k\sigma (E+Q_{E})-rI_{d}-\mu I_{d} \nonumber \\
\dot{R_{u}}&=&rI_{u}-\lambda\dot{I_{d}}R_{u} \label{One_Model_01}\\
\dot{R_{d}}&=&rI_{d} \nonumber \\
\dot{D}&=&\mu I_{d} \nonumber\\
\dot{Q_{S}}&=&\lambda\dot{I_{d}}S \nonumber\\
\dot{Q_{E}}&=&\lambda\dot{I_{d}}E-\sigma Q_{E} \nonumber\\
\dot{Q_{u}}&=&\lambda\dot{I_{d}}(I_{u}+R_{u})+(1-k)\sigma Q_{E}, \nonumber
\end{eqnarray}
subject to the constraint, 
\begin{equation}
S+E+I_{u}+I_{d}+R_{u}+R_{d}+D+Q_{S}+Q_{E}+Q_{u}=N.
\end{equation}
Here, the state variables represent the compartments as described (and labeled) above. The constraint equation ensures that all the compartments add up to the population size $N$. In the context of social distancing, the dynamics of the overall social distancing class is given by,
\begin{equation}
\dot{Q}= \lambda\dot{I_{d}}(S+E+I_{u}+R_{u})-k\sigma Q_{E}.
\end{equation} 

\subsubsection{The Effect of Migration and Regional Spread}
\label{Subsection_Migration}

One of the measures taken to combat the spread of COVID-19, both globally, as well as in India, has been travel restrictions. Such policies came from the apprehension that infectious patients traveling from different countries/states will contribute to the growth of the epidemic in the destination. Accordingly, we develop a model that allows for travel in between different regions. The model also helps us gain insights about how much travel and migration contributes to the growth of cases in a region and if and when travel restrictions should be put in place. Consider a matrix $J_{n\times n}$ that holds information about how many individuals travel from each region $i$ to another place $j$, per capita of the source region,
\begin{equation}
J=\begin{bmatrix}
0&J_{12}&\dots &J_{1n}\\
J_{21}&0&\dots &J_{2n}\\
\vdots&\vdots&\ddots&\vdots\\
J_{n1}&J_{n2}&\dots &0\\
\end{bmatrix}
\end{equation}
Therefore, if the population of the region $i$ is $P_i$, then we can find an expression for this population of the region from the equation,
\begin{equation}
\label{Eq-Diffusion}
\frac{dP_{i}}{dt}=\eta(P_{i},t)+\sum_{j\ne i}J_{ji}P_{j}-\sum_{j\ne i}J_{ij}P_{i}.
\end{equation}
This is the generalized form of the diffusion model proposed in \cite{Barthelemy2019} for the growth of cities, motivated by a similar model proposed in the context of wealth distribution \cite{Bouchaud2000}. Here, the first term accounts for the population growth due to domestic reproduction, the second term accounts for in-migration or immigration and the third term accounts for out-migration or emigration. This model can be built on to explain the dynamics of the spread of infection on a network of connected regions. Therefore, from here, we can find the corresponding equations of $S_{i}$, $E_{i}$, $I_{u,i}$ and $I_{d,i}$ from the following ODEs,
\begin{eqnarray}
\label{Eq_Model_Mig}
\dot{S_{i}}&=& -(\beta_{u,i}I_{u,i}+\beta_{d,i}I_{d,i})S_{i}-\lambda_i\dot{I_{d,i}}S_{i}+\sum_{j\ne i}J_{ji}S_{j}-\sum_{j\ne i}J_{ij}S_{i}\nonumber\\
\dot{E_{i}}&=&(\beta_{u,i}I_{u,i}+\beta_{d,i}I_{d,i})S_{i}-\sigma E_{i} -\lambda_{i}\dot{I_{d,i}}E_{i}+\sum_{j\ne i}J_{ji}E_{j}-\sum_{j\ne i}J_{ij}E_{i}\nonumber\\
\dot{I_{u,i}}&=&(1-k)\sigma E_{i}-rI_{u,i}-\lambda_{i}\dot{I_{d,i}}I_{u,i}+\sum_{j\ne i}J_{ji}I_{u,j}-\sum_{j\ne i}J_{ij}I_{u,i}\nonumber\\
\dot{I_{d,i}}&=&k\sigma (E_{i}+Q_{Ei})-rI_{d,i}-\mu I_{d,i}+\sum_{j\ne i}J_{ji}I_{d,j}-\sum_{j\ne i}J_{ij}I_{d,i}
\end{eqnarray}
Here, the penultimate terms represent a migration of individuals into the region, while the final terms represent the migration of individuals out of the region. Note that, the differential equations describing the change in the other classes are the same as that in (\ref{One_Model_01}). Additionally, even recovered individuals can travel but their travel does not have any implications, as they are neither infected not susceptible and cannot contribute to the rise of cases in the sink region, either at present or in the future. Individuals who are social distancing are assumed to not be traveling. Equations (\ref{Eq_Model_Mig}) are a model for the general case with no travel restrictions. Elements of the matrix $J$ can be used to account for the volume of travel. The only assumption made here is that the compartments are well-mixed in the population and that the probability of an individual traveling is independent of the compartment they belong to. Therefore, the set of travelers is assumed to be a random sample of the population of the source region.

The model can further be tweaked based on different policies and guidelines. For instance, it is a common practice to allow only individuals without symptoms or only individuals with negative COVID-19 test results to travel/enter a state/country. Such modifications can be introduced into the model to disallow individuals in the $I_d$ (and possibly $I_u$) class from traveling, leading to a modified system of equations similar to (\ref{Eq_Model_Mig}) but without the last two terms in the equation for $\dot{I_{d,i}}$ (and $\dot{I_{u,i}})$. By studying the ratio of daily infections caused by migration to the total number of infections, we can judge whether the contribution of migration and travel is significant enough to employ travel restrictions. Furthermore, a total estimate of cases caused due to migration can be made by summing over all cases that may have been caused due to migration and travel in time.

In order to quantify the contribution of an excess number of travelers $\sum_{j\ne i}\Delta J_{ji}E_{j}$, we need to study the ratio of its contribution to the number of (already) exposed individuals in the region, \textit{i.e.,} if $E_{i}$ is the number of exposures in region $i$ without the new set of travelers, we need to study the following ratio:
\begin{equation}
\rho_{i}=\frac{\sum_{j\ne i}\Delta J_{ji}E_{j}}{E_{i}}=\sum_{j\ne i}\Delta J_{ji}\frac{E_{j}}{E_{i}}.
\end{equation}
We call this the migration coefficient. If the migration coefficient is small, it means that the contribution of $\Delta j$ to more travelers to the number of exposures is negligible, and that the surge in cases is due to the domestic growth of cases. This is a useful measure for policy makers as they can use the migration coefficient to decide if open borders would cause a significant increase in the number of cases. An interesting observation we can make is that this ratio does not depend on the absolute number of exposures in the region or in other regions. Rather, it depends on the ratio of cases of the state of origin of migration ($j$) to the destination state ($i$), and the rate of migration.

\subsubsection{The effect of finite Healthcare System Capacity}
\label{Subsection_HCsystem}

In Subsection \ref{Subsection_SEIRQ_model}, it has been assumed that the death rate is a constant, making the instantaneous change in the number of deaths a linear function of the number of active infections, in equation (\ref{One_Model_01}). This is a common assumption in compartmentalized models, and in most cases, fairly reasonable. The approximation works as long as the following criteria are met. 
\begin{enumerate}[(i)]
\item The variant does not mutate or change in potency to become more or less fatal.
\item Medical science, technology or infrastructure does not improve significantly enough to bring down the number of deaths per hospitalization. 
\item The fraction of the cases that get medical attention of the total number of cases which require medical attention, stays constant. 
\end{enumerate}
Failing any one of these, the death rate cannot be modeled as a constant. Here, we focus our attention specifically to the third criterion. We do so because both the first and the second criteria, while holding great relevance to policy, are difficult to model. While it is possible to estimate the form of the function that describes the combined effect of changes in the advances in treatment and the potency of the virus, it is difficult to separate the two out, without more microscopic data and suitable explanatory variables. Deviation from the third criterion too has great consequences to public policy.

Ever since the beginning of the pandemic, there have been lots of conversations about the effectiveness of lockdowns and their necessity (or lack thereof). Popular consensus says that lockdowns reduce the rate of transmission of the infection, but also delay the arrival of the peak and the subsequent fall. Therefore, we do not know if lockdowns in general reduce the number of deaths. Moreover, lockdowns bring with them several economic consequences. Nevertheless, there are two circumstances in which lockdowns may be useful, namely, to temporarily delay the peak so that hospitals can be better prepared, or to avoid the number of serious cases going higher than the capacity of the healthcare system. Proper medical treatment can reduce the mortality rate of COVID-19 and other similar diseases. Therefore, if the number of cases requiring medical care exceeds the number of available hospital beds, the mortality rate will increase. We propose a model to calculate the maximum load that a healthcare system can take, and use it to guide policy makers to determine when to put a lockdown in place. Furthermore, the model lets us quantify the role of medical treatment in reducing the mortality rate of the epidemic. In order to model this, we suppose that a fraction $\xi$ of active infections $I_{d}$ require hospitalization, the maximum capacity of the healthcare system is $\xi K$, the mortality rate with hospitalization is $\mu$ and the mortality rate without hospitalization is $\nu$. If the number of active infections $I_{d}$ is less than $K$, then the relationship is given by equation (\ref{One_Model_01}).
\begin{equation*}
\dot{D}=\mu I_{d}.
\end{equation*}
However, if the number of active infections exceed the number $K$, then the number of individuals who need hospitalization will be more than $\xi K$, then there will be $\xi (I_{d}-K)$ individuals who need hospitalization but will not get any and therefore the the number of deaths without hospitalization will be proportional to ($I_{d}-K$) with rate $\nu$. Therefore, if $I_{d}>K$, then we have, 
\begin{equation}
\dot{D}=\mu K+\nu(I_{d}-K).
\end{equation}
Therefore, we can write the general function as,
\begin{equation*}
\dot{D}=\begin{cases}
\mu I_{d},~I_{d}\le K,\\\mu K + \nu (I_{d}-K),~\text{otherwise},
\end{cases}
\end{equation*} 
or 
\begin{equation}
\dot{D}=\mu \min\{I_{d},K\}+\nu \max\{I_{d}-K,0\}
\end{equation}
with $\nu - \mu$ being the difference in the death rates. Accordingly, we can find the number of lives saved $\mathcal{L}$ by the healthcare infrastructure between any two times $t_{a}$ and $t_{b}$ as,
\begin{equation}
\mathcal{L}=\int\limits_{t_{a}}^{t_{b}}(\nu-\mu)I_{d}dt.
\end{equation}

\subsection{Parameter Estimation}
\label{Subsection_Estimation_of_Unknown_Parameters}
	
For estimating the unknown parameters, we use a least squares approach, where the loss function is given as the square of the residuals. If the data is given by $d_{1},d_{2},\dots,d_{n}$ and the prediction of the given parameters are $Y_{1},Y_{2},\dots,Y_{n}$, then the loss function is given by,
\begin{equation}
\mathcal{LF}=\sum\limits_{i=1}^{n}(d_{i}-Y_{i})^{2}.
\end{equation}
For the estimation of the fracture of the healthcare system and the multiple death rates, we use a 3-parameter gradient descent, parametrizing not only the 2 death rates but also the maximum capacity $K$. However, for the estimation of the parameters of the compartmentalized SEIRQ model, the predicted curves cannot be found analytically, and needs to be determined numerically, and the derivative of the loss function, with respect to the parameters, does not have an analytical form. Therefore, we execute our least squares optimization using Python's \texttt{Scipy.integrate} and \texttt{Scipy.optimize} packages. We first use the \texttt{odeint} function to solve the system of ODEs, and define the objective functional for the problem. The solver numerically solves the system of ODEs. Therefore for a parameter vector $\Theta$, we have a solution of equation (\ref{One_Model_01}) as a set of prediction curves for the different state variables. To get a good estimate of $\Theta$ from the data, we would like to minimize the loss function as defined above. Therefore we aim to find $\Theta$ such that the solution of the system of ODEs is most compatible with the data we have. Since we have three sets of data in the confirmed cases, recovered cases and deaths, we would want a loss function minimized such that the total discrepancy between the solution of the ODEs and the data is minimum. For this we rewrite our dataset to be a concatenated array of confirmed cases, recoveries and deaths. We then use the \texttt{curve\_fit} function to find the parameters that fit the data the best for that objective function. The \texttt{curve\_fit} function works on the objective function which returns a concatenated array of the solution of the system of the ODEs in confirmed cases, recoveries and deaths. For the estimation, we specify reasonable upper and lower bounds for the parameters. We make guesses about the bounds of the initial conditions based on the data available, and we set the bounds for $\lambda$ such that none of the compartments attain negative occupancies in the duration of the study. Then the \texttt{curve\_fit} function returns the optimum set of parameters.

\subsection{Error Analysis}
\label{Subsection_Error_Analysis}
	
To get a measure of the error in our estimation of the parameters, we plot the training error and calculate the $95\%$ confidence intervals for each of the estimated unknown parameters. For the training error, we plot the absolute difference between the data and the simulation for the estimated parameters, and scale the error by the value of the data point. We individually look at the relative training errors in each of the data for confirmed cases, recovered cases and deaths. We also calculate confidence intervals by a method called parametric bootstrap \cite{Efron1986}. In particular, we use a method called residual bootstrap. A residual bootstrap is simply a method of repeated sampling (with replacement) of the residues obtained at each data point (training error), and using this to generate a synthetic distribution for the population, where each residue $r_i$ is defined as,
\begin{equation}
r_{i}=d_{i}-Y_{i}.
\end{equation}
In this method, we simulate the model with the estimated parameters and calculate the residual vector. Here, we construct the residual vector by concatenating the sets of residuals left in each of the time series of confirmed cases, reported recoveries and deaths. We assume the set of residuals to represent the distribution of residuals and generate $1000$ new time series wherein the synthetic (bootstrapped) data is generated by adding synthetic residuals randomly sampled (with replacement) from the real set of residuals to the predicted data (from the model). For each of these synthetic data sets, we estimate the parameters in the method described above. This gives us $1000$ estimated values for each parameter after bootstrapping. This set of parameters is assumed to be the distribution for the population, and the $95\%$ confidence interval for each parameter is estimated from them by simply considering the data points at the $2.5$-th and the $97.5$-th percentiles. 
	
\section{Results in the Context of India}
\label{Results}

In this Section, we present the detailed results of the extensive analysis carried out on the basis of the methodologies outlined in the preceding Section. The findings are based on the actual data of the spread of COVID-19 in India. In the same order as the presentation of the methodology, we begin with the quantification of the social distancing in the context of the outbreak in India. This was followed by a more localized study of the regional spread of COVID-19 in India, across the states and union territories. Finally, we carry out the quantification of the capacity and role of the healthcare system, especially the determination of the breaking point of the system.

\subsection{Quantifying Social Distancing}
We first estimate the parameters of the model and use it to gain inferences about the spread of the pandemic in the country. We also show that the model fits the data well, and analyze the errors, as described in the previous Section. As the social distancing class is unique about our model, we also use the estimated parameters to find the number of individuals moving in and out of social distancing for the period considered.
\begin{figure}[h]
\begin{subfigure}{.5\textwidth}
\centering
\includegraphics[width=1\linewidth]{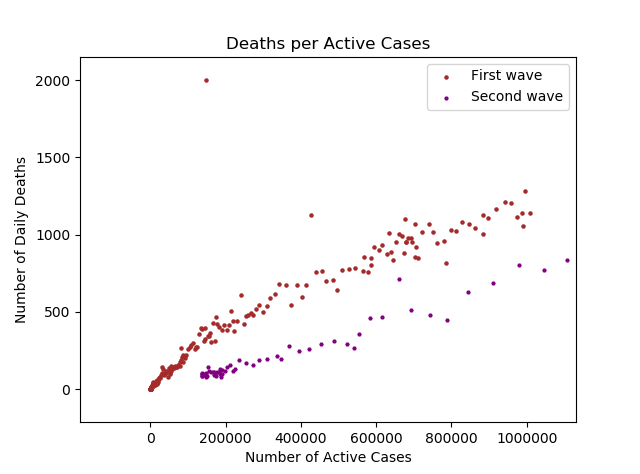}
\caption{Variation of death rates in first and second wave (pre-peak)}
\label{fig:deaths-per-active-1v2}
\end{subfigure}
\begin{subfigure}{.5\textwidth}
\centering
\includegraphics[width=1\linewidth]{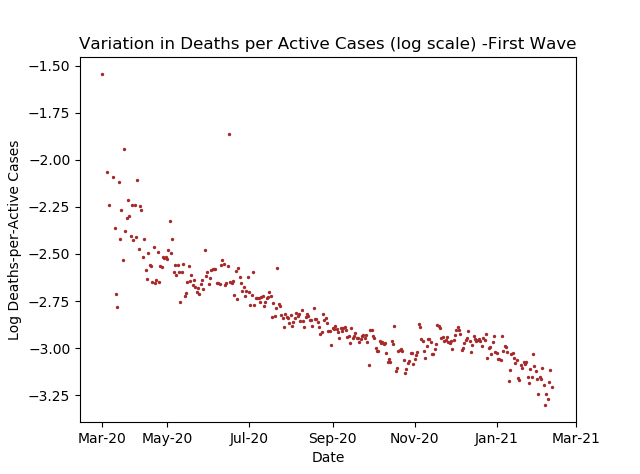}
\caption{Variation of death rate in first wave (log scale)}
\label{fig:log-death-rate---first-wave}
\end{subfigure}
\caption{(a) \textbf{Variation between waves}: The figure shows how the number of daily deaths per active case were different between the first wave and second wave in India (b) \textbf{Variation within wave}: There is a clear decline in the death rate in the first wave.}
\label{fig:Scatters DR}
\end{figure}
Our model has $10$ compartments and therefore 10 state variables, out of which $3$ variables, namely, $I_{d}$, $R_{d}$ and $D$ are the ones for which the data was available. Therefore, we also have information on the initial conditions of these variables. However, the initial conditions of the remaining variables needs to be estimated for the data. This estimation will give more insights on several variables, information on which cannot be found from existing data. Moreover, there are $7$ parameters that need to be estimated from the data. These parameters give us information about the spread of the pandemic.
\begin{figure}[h]
\begin{subfigure}{.33\textwidth}
\centering
\includegraphics[width=1\linewidth]{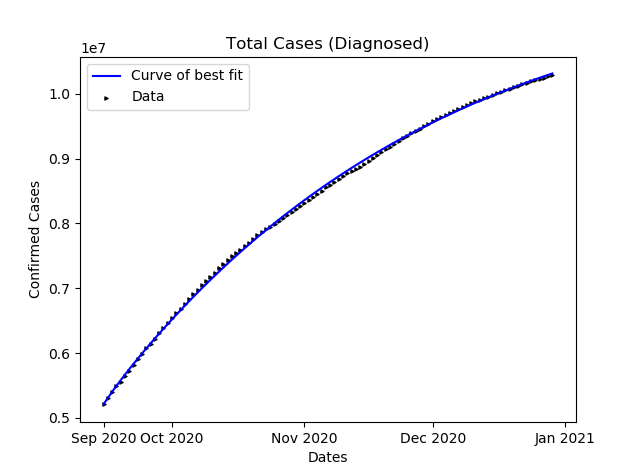}
\caption{ Total Confirmed Cases}
\label{fig:confirmed-cases---diagnosed---fit}
\end{subfigure}
\begin{subfigure}{.33\textwidth}
\centering
\includegraphics[width=1\linewidth]{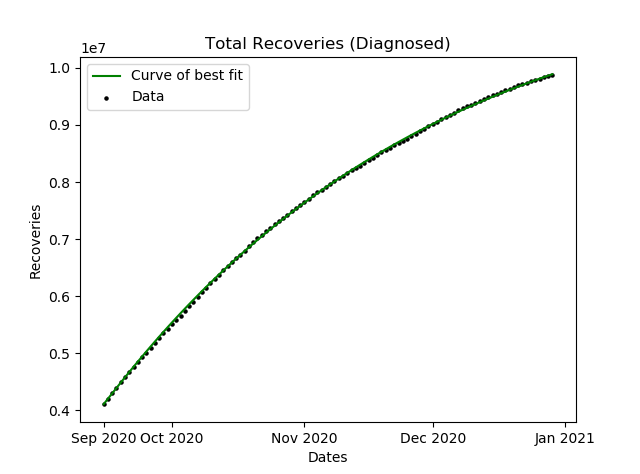}
\caption{Total Recoveries (Diagnosed)}
\label{fig:recovered-cases---diagnosed---fit}
\end{subfigure}
\begin{subfigure}{.33\textwidth}
\centering
\includegraphics[width=1\linewidth]{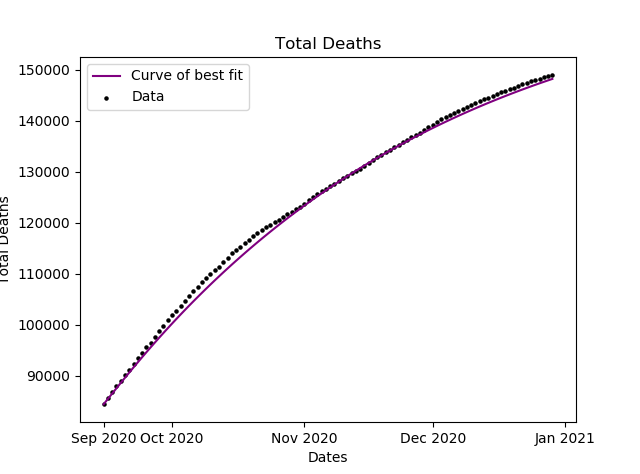}
\caption{Total Deaths}
\label{fig:deaths---diagnosed---fit}
\end{subfigure}
\newline
\begin{subfigure}{.33\textwidth}
\centering
\includegraphics[width=1\linewidth]{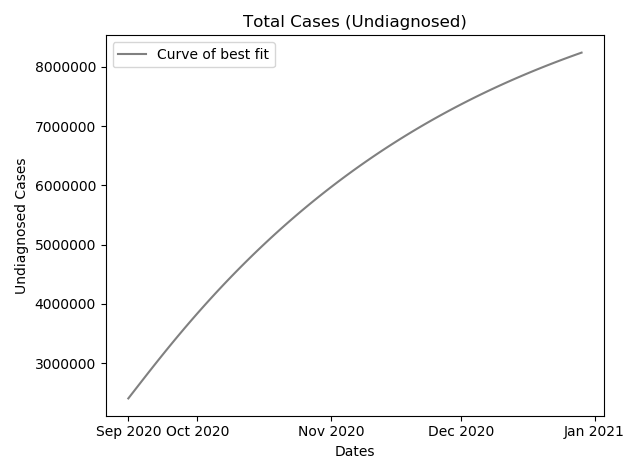}
\caption{Total Undiagnosed Cases}
\label{fig:cases---undiagnosed---fit}
\end{subfigure}
\begin{subfigure}{.33\textwidth}
\centering
\includegraphics[width=0.95\linewidth]{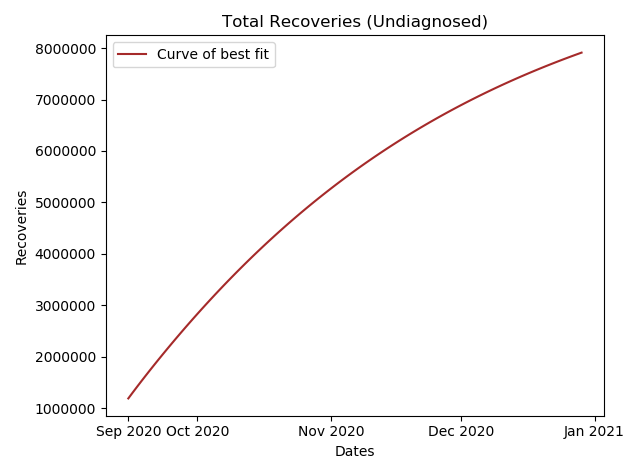}
\caption{Total Recoveries (Undiagnosed)}
\label{fig:recovered-cases---undiagnosed---fit}
\end{subfigure}
\begin{subfigure}{.33\textwidth}	\centering
\includegraphics[width=1\linewidth]{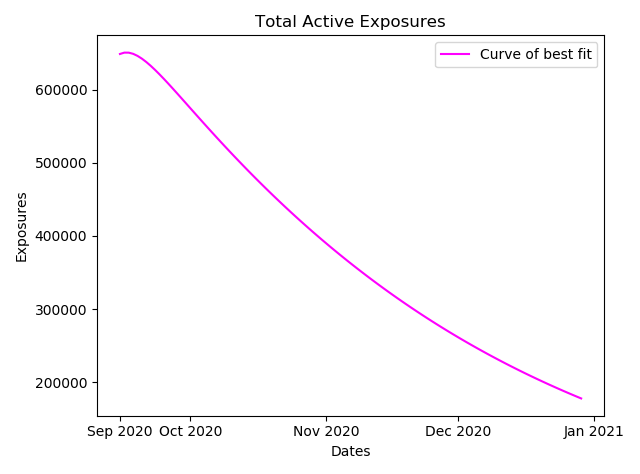}
\caption{Active Exposures}
\label{fig:exposed---fit}
\end{subfigure}
\caption{\textbf{The Curves of Best Fit as determined by the Parameter Estimation}. Here the curves are determined so that they explain the $3$ data sets (Confirmed Cases, Recoveries and Deaths) better together, and not independently. The parameters are determined such that the aggregate least squares error of the $3$ data curves is minimized.}
\label{fig:fig4}
\end{figure}
	
Although the data for India is available from March 2020 to October 2021, our model works better for shorter periods of time. This is because our model hinges on constant parameters, while in reality, parameters are anything but constant. For instance, India is thought to have been impacted by three main variants, namely, the original COVID-19 virus, the Delta variant and the Omicron variant. Each of these variants brought with them waves of infections. Since our parameters are related to actual measurable quantities like average incubation time, death rate etc. it would be unreasonable to assume that these values remain constant throughout different variants. For instance, we can see that the death rate in the first wave and the second wave on its way to the peak were very different, with the death rate in the first wave being significantly higher.

Furthermore, variation is not only due to variants but also due to improvements in medical knowledge and treatment facilities. If we consider the first wave alone, we can see that the death rate reduces gradually overtime. This decrease is particularly pronounced in the first few months and last few months of the pandemic with September-December showing some stability, although November shows a dip in the death rate, which soon rectifies itself. There is no reason why one should expect death rate to be the only parameter which varies with time and therefore, it is better to restrict our analysis to a short span (a few months) so that assuming the parameters are constant do not cause serious deviations from reality. This is also why we specifically refrain from making predictions about the pandemic into the future as we recognize that the value of the parameters are unlikely to remain constant.
\begin{table}[h]
\begin{center}
\begin{tabular}{|c|c|c|}
				\hline
				Estimated Parameters & Estimated value & 95\% confidence intervals \\      
				\hline
				$\lambda$ &  $8.8920\times 10^{-5}$ & $[6.4462\times 10^{-9},\ 1.0000\times10^{-4}]$\\
				\hline
				$\beta_{u}$ & $4.9263\times 10^{-1}$ & $[3.7836\times10^{-1},\ 7.0066\times10^{-1}]$\\
				\hline
				$\beta_{d}$ & $ 4.0642\times 10^{-1}$ & $[2.0218\times 10^{-1},\ 5.5209\times 10^{-1}]$\\
				\hline
				$k$ & $4.6298\times 10^{-1}$ & $ [2.9467\times 10^{-1},\ 5.2058\times 10^{-1}]$\\
				\hline
				$\sigma$ &$2.6808\times 10^{-1}$ & $[2.5000\times 10^{-1},\ 4.5851\times 10^{-1}] $\\
				\hline
				$r$&$9.2898\times 10^{-2}$&$[9.1596\times 10^{-2}, 1.4852\times 10^{-1}]$
				\\
				\hline 
				$\mu$&$1.0250\times10^{-3}$& $[1.0250\times10^{-3}, 3.3729\times 10^{-2}]$\\
				\hline
				$E(0)$ & $648657$ & $ [238192,\ 883282]$\\
				\hline
				$I_{u}(0)$ & $ 1220865$ & $ [871917,\ 1995837]
				$\\
				\hline
				$R_{u}(0)$ & $1186785$ & $  [839130,\ 1382516]
				$\\
				\hline 
				$Q_{s}(0)$ & $1120405558$ & $[1022027510,\ 1175406760]
				$\\
				\hline 
				$Q_{e}(0)$ & $158529$ & $ [50002,\ 499891]$\\
				\hline
				$Q_{u}(0)$ & $2934535$ & $ [2049996,\ 3429924]$\\
				\hline
\end{tabular}
\caption{\label{Table_01} Estimation of Unknown Parameters.}
\end{center}
\end{table}

For the estimation of the national-level parameters, we consider the period between 17th September 2020 to 31st December 2020. The choice of this period is not arbitrary and is a natural best choice. This is for several reasons. Firstly, we choose the first wave of infections where the number of cases was significantly lower than those in the second wave, which facilitates better testing and therefore more robust data. Furthermore, our model considers only deaths from detected infections. This is because it is assumed that any serious infection that could potentially have been fatal will require medical attention and will therefore have been diagnosed. This hypothesis will not stand in the second wave where it was widely reported that the number of cases may have been more than what the healthcare infrastructure could handle, and therefore it is not possible to ascertain if every COVID-19 related death would have followed a COVID-19 diagnosis. Furthermore, even within the first wave, not all periods have data that satisfies the assumptions of our model. For instance, our model does not assume a vaccination compartment, \textit{i.e.,} our model works best in a population with no vaccination or full vaccination. Since India undertook an aggressive vaccination campaign in 2021, a model that considered a variable vaccination state would better explain the data. In India getting vaccinated was only restricted to healthcare workers in 2020, which is an extremely small number, almost negligible in comparison to the total population. Furthermore, the first few months of the spread of COVID-19 had constantly changing norms as India first had over 2 months of a stringent lockdown, and the relaxations called ``unlockdowns'' came in phases rather than all at once, and therefore the period of the first few unlockdowns had constantly changing norms every two weeks which would break our constant parameter assumptions. Furthermore, we can see that there is a lot more variation in deaths per active case in the first few months. This is because the number of cases itself was small, much smaller than the value at the peak. Thus, small errors on an absolute scale would be very large errors in the first few months of the spread and would make the estimation less insightful. There are also discrepancies in the data in the earlier period, for instance, on 16th June 2020, there was a sudden spike in the deaths (greater than 2000). These problems do not persist in the latter part of 2020. Another advantage to using data from 2020 is the existence of travel data in 2020 and previous years which let us quantify the effect of fall in tourism (a proxy for travel) in reducing the spread of COVID across states, which we study in subsequent subsections.

\begin{figure}[h!]
\begin{subfigure}{.5\textwidth}
\centering
\includegraphics[width=1\linewidth]{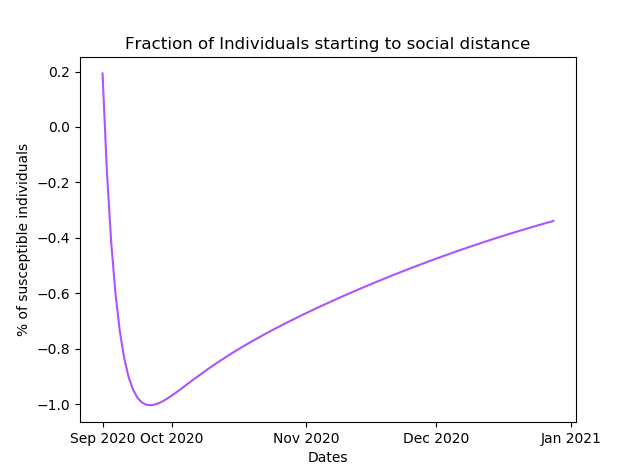}
\caption{Fraction of susceptible individuals moving in/out of social distancing}
\label{fig:frac-sd}
\end{subfigure}
\begin{subfigure}{.5\textwidth}
\centering
\includegraphics[width=1\linewidth]{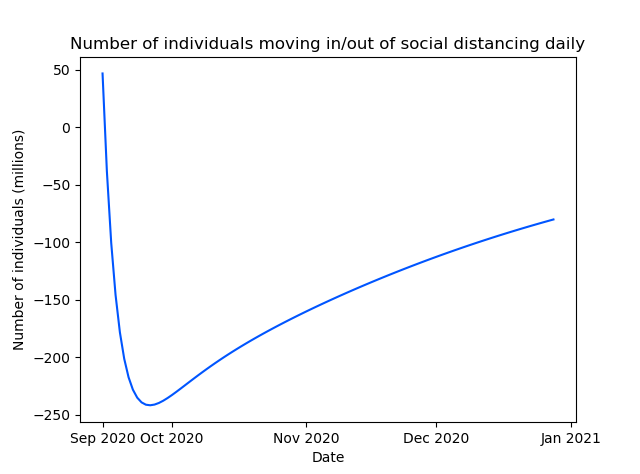}
\caption{Total number of individuals moving in/out of social distancing}
\label{fig:numb-sd}
\end{subfigure}
\caption{\textbf{Individuals moving in and out of the social distancing class daily}: (a) Fraction of susceptible cases (b) Total number of individuals. A positive (negative) value means more (less) individuals are beginning to social distance. }
\label{fig:SD}
\end{figure}

Table \ref{Table_01} gives the result of the parameter estimation for the $6$ initial conditions and $7$ constants of the system of differential equations (\ref{One_Model_01}). The $7$-th initial condition $S(0)$ is estimated from the constraint equation. The confidence intervals are found using the residual parametric bootstrap method as described above. Based on the estimated parameters, we can observe the fraction and absolute numbers of individuals moving in (positive rate) and out (negative rate) of the social distancing class, as seen in Figure \ref{fig:SD}.
We can observe from Figure \ref{fig:SD} that initially when the active cases were at the peak, the rate of social distancing was positive, with more individuals moving into social distancing everyday. However, as the number of cases began falling, the rate of social distancing fell sharply, with individuals beginning to come out of social distancing and returning to ``normalcy''. As the rate of fall of cases, eventually reduced, the rate at which individuals leaving social distancing also reduced. This is because our model assumes the rate of social distancing to be a product of the rate of change of diagnosed infections and the number of susceptible individuals. In the period considered the number of susceptible individuals did not change by as large an amount as the number of daily infections and thus the trends are largely dominated by $\dot{I_{d}}$.
	
Table \ref{Table_01} also gives us insights about the behavior of other parameters. For instance, we find that $\beta_{u}$ is greater than $\beta_d$. This means undetected infections contribute to new cases at a faster rate than detected ones. This is consistent with intuitive expectations as individuals who have been diagnosed for the infection may isolate after the diagnosis, while those who have not been diagnosed will not take any additional precautions. Furthermore, we find that more than half of the total cases go undiagnosed. This is possibly because a large number of COVID-19 infections are mild and asymptomatic, do not require medical attention, leading individuals to dismiss it as a common cold and not get tested. We also find that the average incubation time of the virus is $\displaystyle{\frac{1}{\sigma}}$ or $3.73$ days, after which the exposed individuals become infectious.
	
\begin{figure}[h]
\centering
\includegraphics[width=0.5\linewidth]{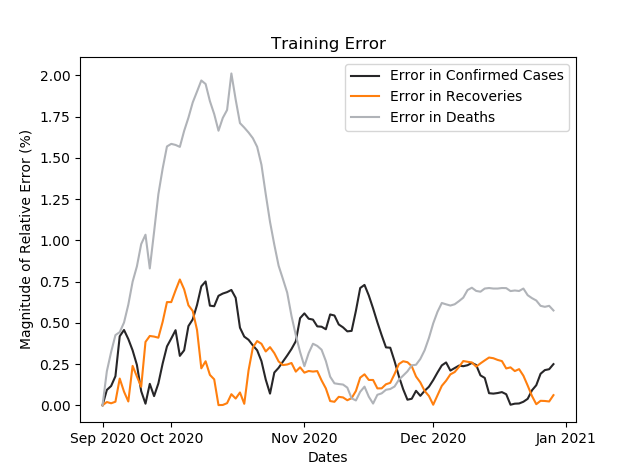}
\caption{\textbf{The Training Error relative to the data}}
\label{fig:training-error}
\end{figure}

The training error of the parameter estimation can be found in Figure \ref{fig:training-error}. One can see that the error is typically less than $1\%$, with error in the number of deaths going slightly higher (but still $<2\%$). This higher relative error in the estimation of the death rate is due to the number of deaths being significantly smaller than the total number of confirmed cases or recoveries, thus making the relative errors more pronounced due to a smaller scaling factor.  
	
\subsection{The Regional Spread of COVID-19 in India}

Compartmentalized models typically assume populations to be uniform and well mixed, which need not be true. It is entirely possible that the number of cases in different geographical regions are different. Accordingly, we study the skew of the distribution of COVID-19 infections in India, across states, by employing the Lorenz Curve and Gini Coefficient. The Lorenz curve is a graphical representation of the fraction of cases (in the vertical axis) in the bottom $x$-fraction of states (in terms of number of cases). Therefore, the horizontal axis has the bottom $x$ percentile of States from $0$ to $100$, and the vertical axis has the per-capita cases held by that fraction. If the distribution is perfectly equal, then there will be a straight line with slope $1$, which implies that the cases are distributed by population uniformly. The further away the curve is from the line of perfect equality, the more the distribution of cases are skewed, \textit{i.e.,} the cases are concentrated in some states, while other states have fewer cases per capita. The Gini coefficient is a measure of this inequality as it is defined as the area between the Lorenz curve and the line of perfect equality, and therefore quantifies the deviation from the uniform distribution. A Gini coefficient of $0$ implies an equal distribution and a Gini coefficient of $1$ implies all the cases are in one state.

\begin{figure}[h]
\begin{subfigure}{.33\textwidth}
\centering
\includegraphics[width=1\linewidth]{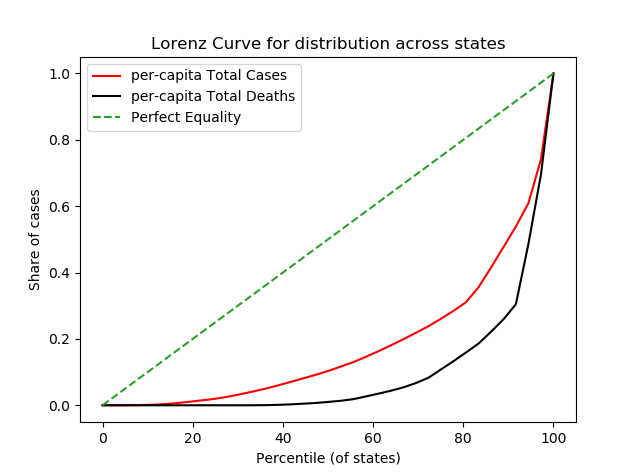}
\caption{1st June 2020}
\label{fig:lorenz-june20}
\end{subfigure}
\begin{subfigure}{.33\textwidth}
\centering
\includegraphics[width=1\linewidth]{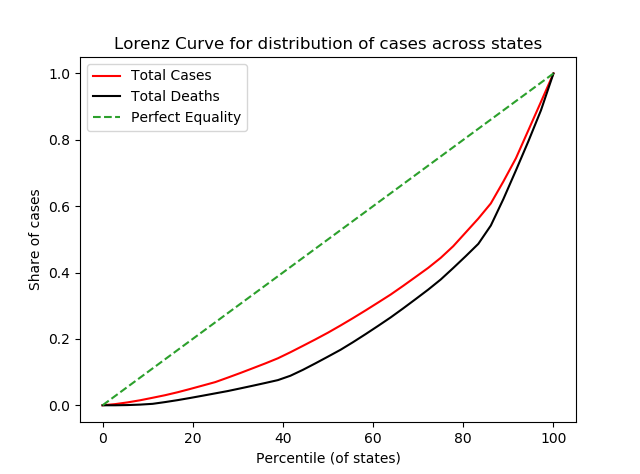}
\caption{1st February 2021}
\label{fig:lorenz-feb21}
\end{subfigure}
\begin{subfigure}{.33\textwidth}
\centering
\includegraphics[width=1\linewidth]{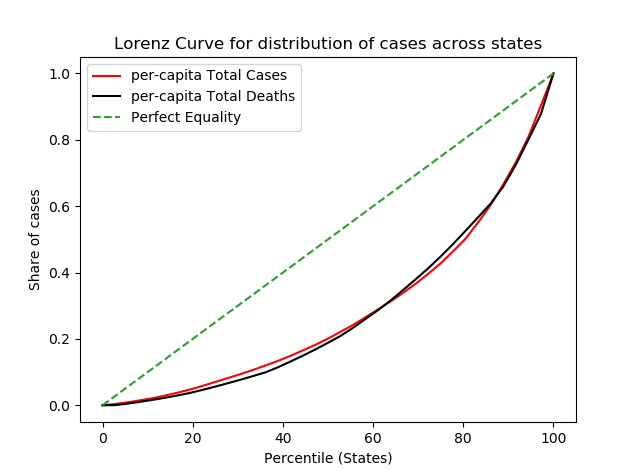}
\caption{31st October 2021}
\label{fig:lorenz-end}
\end{subfigure}
\caption{\textbf{Lorenz Curves of Distribution of Total Cases and Total Deaths}: (a) 1st June 2020 (b)1st February 2021 (c) 31st October 2021}
\label{fig:Lorenz}
\end{figure}
We plot the Lorenz Curves at three different times, namely, at the end of the lockdown on 1st June 2020, around the end of the first wave on 1st February 2021 and around the end of the second wave on 31st October (where our dataset ends). We can see from the Lorenz curves that even though the distribution was very unequal in June 2020, immediately after the lockdown, however, over time the spread of cases becomes more evenly distributed between states. We also compute the Gini coefficient and study how that varies over time. We compute the Gini coefficient of total confirmed cases, total deaths and active cases.
\begin{figure}[h!]
\begin{subfigure}{.5\textwidth}
\centering
\includegraphics[width=1\linewidth]{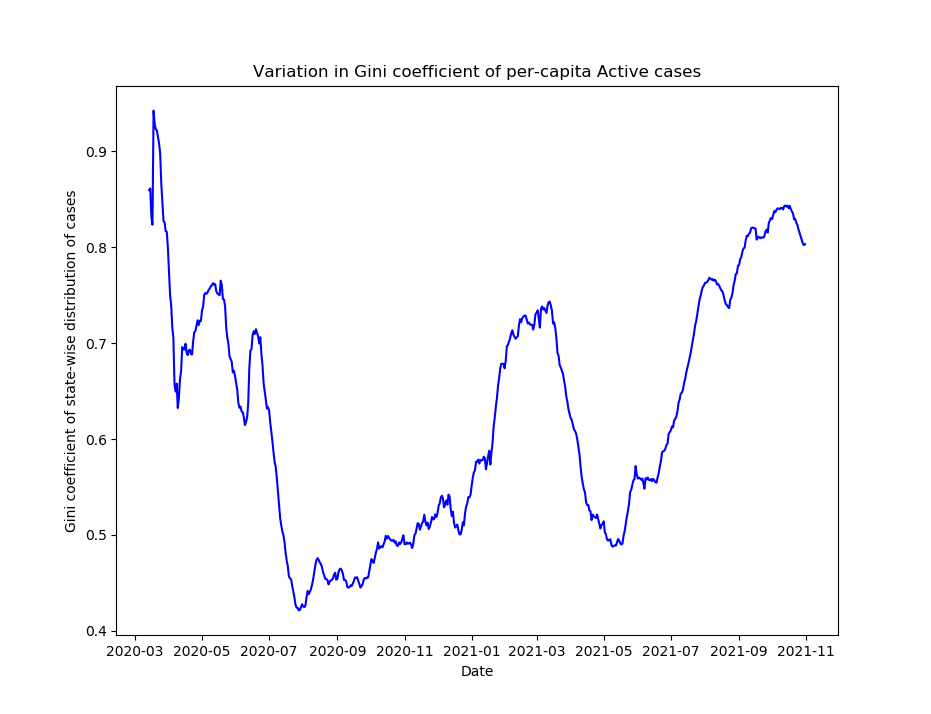}
\caption{Gini - Total Confirmed Cases and Deaths}
\label{fig:gini-case-death}
\end{subfigure}
\begin{subfigure}{.5\textwidth}
\centering
\includegraphics[width=1\linewidth]{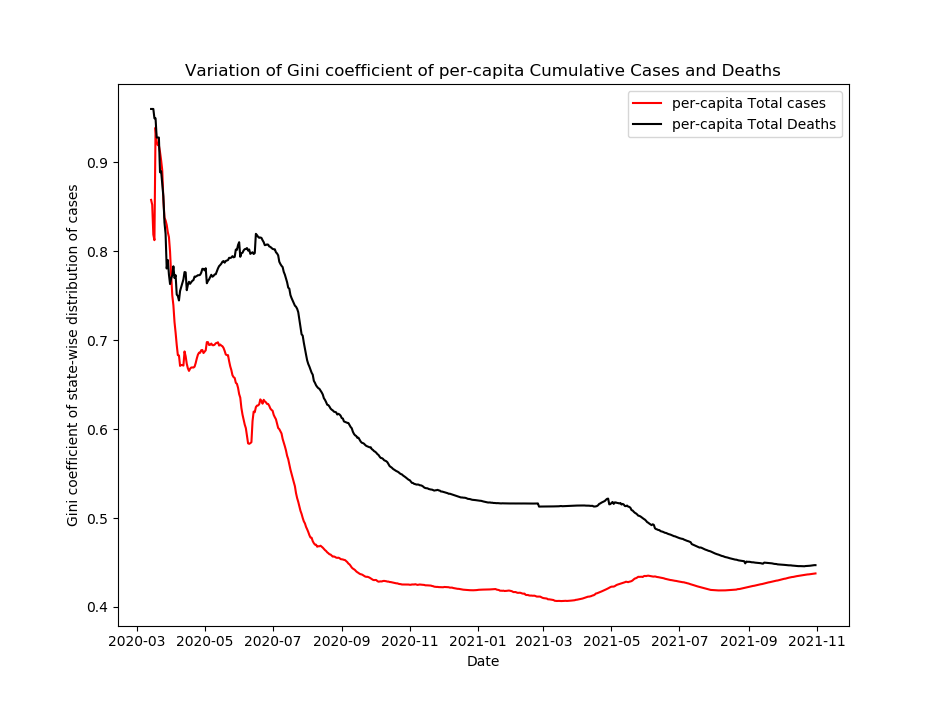}
\caption{Gini - Active Confirmed Cases}
\label{fig:gini-active}
\end{subfigure}
\caption{\textbf{Change of Gini coefficient with time}: The Gini coefficient on any particular date is computed based on the distribution of the numbers across states in India. (a) \textbf{Total Confirmed Cases and Deaths} (b) \textbf{Active Confirmed Cases}}
\label{fig:Gini} 
\end{figure}
From Figure \ref{fig:Lorenz} and Figure \ref{fig:Gini}, we can see that initially, the inequality in the per-capita deaths is even higher than the inequality in the per capita cases, but this too changes over time and eventually the inequality in the per-capita cases and per-capita deaths becomes somewhat equal. This implies that most deaths were initially concentrated to a few states, with some states having a disproportionately large number of deaths per case, but over time, the states which had low death rates initially had more deaths (compared to the average) and the distribution of deaths became more equal. We also observe from Figure \ref{fig:Gini} that the Gini of the active cases doesn't come down consistently, while that of total cases does. This implies that the distribution of active cases did not get more equally distributed over time. Instead, the states which had less cases per-capita before, had more cases per capita at later times, thus allowing the cumulative cases to be ``better'' distributed. Therefore, we can see from the figures that there is strong evidence that the distribution of cases is not well mixed. This provides the motivation for the model that studies a country (or the world) as a network of connected regions each with its own set of rates and parameters, with a model that factors in migration (\ref{Subsection_Migration}). For a thorough analysis of the model, we would require data on interstate migration, so that we could estimate the parameters of the large connected system of $n$ differential equations, where $10\times n$ is the number of regions. However, such data is not easily available and the estimation of parameters is not possible at this time.
	
However, the model allows us to study the effect of travel in the transmission of cases. We calculate the migration coefficient ($\rho$) for exposures as defined in subsection \ref{Subsection_Migration}. The data we have accounts for travel due to reasons of residence, reentry, employment, tourism, business, studies, and other common reasons for travel. Therefore, we deduce that it is a suitable measure for studying interstate migration. For the analysis, we consider the 10 states most frequented by travelers in 2020 (as we have in-migration data for domestic travelers in 2020 for these states only). We compare the number of travelers who visited the states in 2020 with the number of travelers who had visited these states in 2019. We restrict our analysis to what would have happened had the number of travelers in 2020 been at par with the numbers from 2019. Therefore, we are interested only in the deficit in the number of visitors and not in the absolute numbers.
\begin{figure}[h!]
\begin{subfigure}{.5\textwidth}
\centering
\includegraphics[width=1\linewidth]{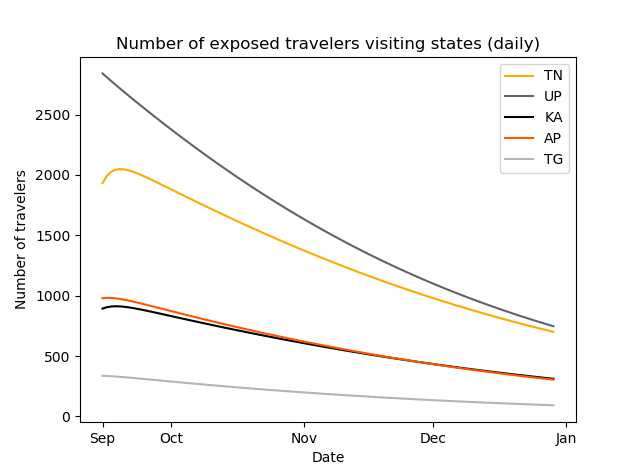}
\caption{The five most frequented states}
\label{fig:exposed-travelers1}
\end{subfigure}
\begin{subfigure}{.5\textwidth}
\centering
\includegraphics[width=1\linewidth]{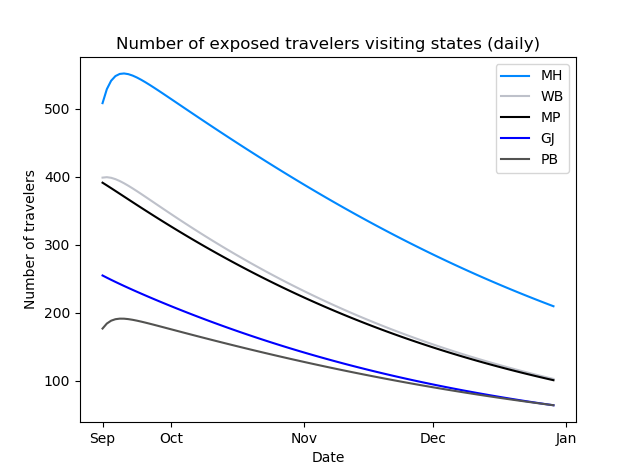}
\caption{The next five most frequented states}
\label{fig:exposed-travelers2}
\end{subfigure}
\caption{\textbf{Number of Exposed Travelers to states (daily)}: (a) Tamil Nadu, Uttar Pradesh, Karnataka, Andhra Pradesh, Telangana (b) Maharashra, West Bengal, Madhya Pradesh, Gujarat, Punjab}
\label{fig:Exposed-Travelers} 
\end{figure}
\begin{figure}[h!]
\begin{subfigure}{.5\textwidth}
\centering
\includegraphics[width=1\linewidth]{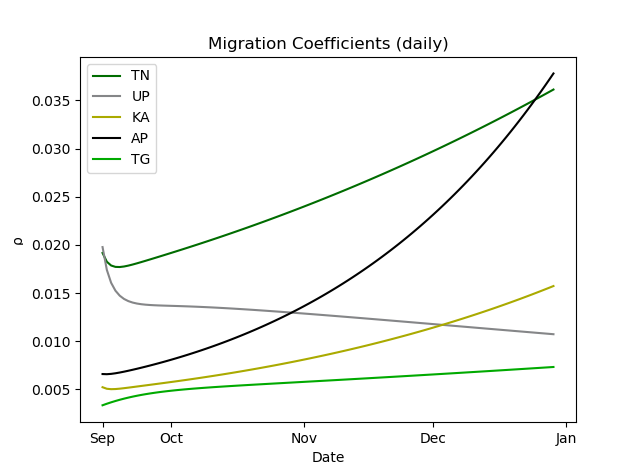}
\caption{The five most frequented states}
\label{fig:mig-coeff-1}
\end{subfigure}
\begin{subfigure}{.5\textwidth}
\centering
\includegraphics[width=1\linewidth]{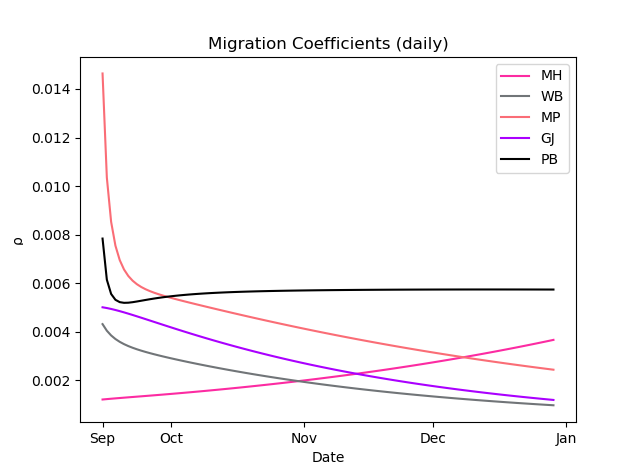}
\caption{The next five most frequented states}
\label{fig:mig-coeff-2}
\end{subfigure}
\caption{\textbf{State-wise Migration coefficients (daily)}: (a) Tamil Nadu, Uttar Pradesh, Karnataka, Andhra Pradesh, Telangana (b) Maharashra, West Bengal, Madhya Pradesh, Gujarat, Punjab}
\label{fig:Mig-Coeff} 
\end{figure}
	
We make the simplifying assumption that the number of travelers who visit each of these states is uniformly distributed across the days of the year. This assumption, although unlikely to hold exactly, helps us in calculations, since we do not have information of the actual seasonal distribution of travelers. We also assume that travel in 2020 went on at the level of 2019, until 23rd March, when the first lockdown was announced, and the remaining numbers are distributed equally over the remaining days of 2020. While considering each state, we divide our system into two components, the focal state and the rest of the country, and we study the effect of migration from the rest of the country to the focal state. Note that since we are only studying the effect on the focal state, we are not interested in out-migration, although infected individuals leaving the state will only make the effect of travel less significant. Also in order to avoid estimating the values of too many parameters at once, as would have been the case if we estimated the number of exposed individuals, with the existing rate of migration, we estimate the number of exposed individuals by estimating parameters in the model without migration. This is because, under our assumptions, the amount of daily domestic migration in India is found to be less than $5\%$ of that of 2019, and thus we assume it to be reasonable on our part to consider that much travel insignificant. We understand that in the process of arriving at an estimate, we have made several simplifying assumptions in this part of the analysis. Therefore, the estimate we arrive at will be a coarse one. Therefore, the results found in this section are intended to be more of a demonstration of the application of the model and we expect to get only a rough estimate of the migration coefficient that is of the correct order, and thus insightful. Here, we can see that migration coefficients are typically very small, mostly less than $0.02$ (with some exceptions) and always less than $0.05$. This tells us that had migration and travel continued at the level of 2019, there would not have been a very large increase in the number of exposures in these states for the period considered.
	
\subsection{Quantifying the capacity and role of the Healthcare system}
	
In previous Subsections, we have discussed in detail why we refrain from testing our model during the second wave, especially in the context of multiple variants, changing parameter values, possibly less robust data and increasing vaccinations (especially in the second half of the year). Therefore, without bothering about the intricacies of the compartmentalized model, we only look at the relationship between active cases and daily deaths, in the period between 11th February 2021 and 7th May 2021. These dates are chosen because they are the dates with the minimum cases and maximum cases in the period between the peaks of the first and second wave, \textit{i.e.,} the time period corresponds to the period when the second wave was on the rise. 
\begin{figure}[h]
\centering
\includegraphics[width=0.48\linewidth]{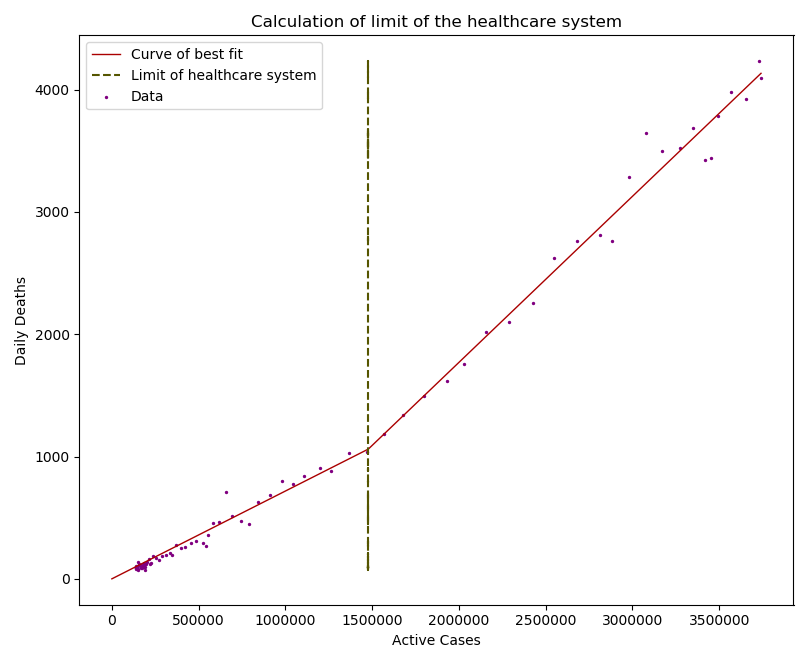}
\caption{\textbf{Breaking Point of the Healthcare system}: We can see a clear change in slope that shows that the death rate changes after the number of infections crosses a threshold.}
\label{fig:hc-limit}
\end{figure}
The reason we can choose to study only the relationship between reported active cases and deaths is because the equation for the number of deaths depends only on active cases and not other variables for which we may not have had actual data.
\begin{table}[h]
\begin{center}
\begin{tabular}{|c|c|}
				\hline
				Parameter & Value\\
				\hline 
				$\mu$ &7.17$\times 10^{-4}$\\
				\hline
				$\nu$ &1.358$\times 10^{-3}$\\
				\hline
				$K$ & 1.4765$\times 10^6$\\
				\hline
			\end{tabular}
\caption{\label{Table_02} Estimation of Maximum Capacity of Healthcare System.}
\end{center}
\end{table}
We observe that the relationship between active cases and daily deaths remains linear throughout but changes slope in between. As stated in subsection \ref{Subsection_HCsystem} this is due to the hospital capacity being pushed to its limit and patients not getting medical treatment. From Figure(\ref{fig:Scatters DR}) we can see that even though the death rate in the first wave was higher and the absolute number of deaths had increased beyond $1000$ deaths, the slope of the daily deaths per active case does not seem to change, when for the second wave, it does. This is because the change of the death rate is actually dependent on the total number of active cases and the active cases in the first wave was always below what the healthcare system could manage.

We also note that vaccinations do not play much of a role as data from \cite{DataIndia} suggests that the number of doses administered by 7th May 2021 was approximately $1\% $of India's population, which means that the number of people who had received two doses were less than $0.5\%$ of India's population, and therefore we do not expect to see much of an effect of vaccination in this period. The death rates as estimated and the maximum possible load of the healthcare system are given in Table \ref{Table_02}. Therefore, we can see that the death rate with hospitalization was $7.17\times 10^{-4}$ and the death rate without hospitalization was $1.357\times10^{-3}$. From this we can see that hospitalization brings down the chances of mortality by $47.2\%$. This analysis also gives us insights about when lockdowns should be implemented, as implementing a lockdown can slow down the rate of infection, thereby arresting the transmission, so that the maximum capacity is never exceeded.
	
\section{Discussion}
\label{Section_Discussion}

The inclusion of the social distancing compartment in the SEIR-based model aims to bridge the gap in the literature between simplified models and real behaviors of populations in a pandemic. Since the extent of social distancing in absolute numbers is not in general a measurable quantity, the model alleviates some of the problem significantly by allowing the estimate of the rate of social distancing from data. While our model assumes a constant rate of social distancing (since we have only trained it with data for a few months), variable rates of social distancing can be studied by dividing the time series into discrete partitions and estimating the extent of social distancing in each of them. Variable rates of social distancing can also be studied by assuming a form of the function of the rate, \textit{i.e.,} instead of the rate being described by a time independent constant, wherein it would be described by a parametric function. Then the parameter estimation would have to be slightly tweaked to estimate more parameters, but the conceptual framework would remain unchanged.

Our modification in the model to allow regional spread was largely motivated by data and policy. We noticed that the distribution of per capita cases and deaths are extremely unequal. We also noted that migration was among the main factors that led to COVID-19 spreading across the world, and major policy decisions today still involve restricting international or interstate travel to curb an influx of cases. Therefore, it was of paramount importance to propose a measure that allows policy makers to determine when it is relevant to restrict travel and close borders, because it is possible that most of the rise in cases of a region is owed to domestic reproduction of cases and not migration. Restricting migration in such a case would be an inconvenience for people and an economic adversity for some industries. We found in the period we considered, that the effect of migration (at the pre-pandemic level) to a rise in exposed cases in a state were rather small for most states, except those with very high rates of immigrants. Our model with migration also has the potential to explain interesting empirical phenomena, like the power-law distribution that was observed between countries towards the beginning of the pandemic when the growth rates were exponential \cite{Blasius2020}. Although we do not investigate this in the article due to the lack of freely available data on international travel and the uncertainty with regard to any real policy implications, the analysis for equation \ref{Eq-Diffusion} under exponential growth and mean-field migration yields a power law \cite{Bouchaud2000}. 

Finally, our model incorporates the healthcare capacity of the system and its effect on the death rate. The second wave that peaked in April-May 2021 in India was devastating in terms of the number of deaths caused. The wave is believed to have been caused by the delta variant. Although the death rate per case is lower in the second wave than the first, the number of people who died in the second wave was much higher, due to the sheer number of cases, which was so large that the healthcare capacity seemed saturated. In our analysis we found two distinct death rates separated by a hard boundary which corresponds to the health care capacity. We found that the death rate without hospitalization is nearly double that with hospitalization, and we found that hospitalization can be crucial in reducing the number of deaths (by $47.2\%$).

Our contribution is largely theoretical, and data has been used to illustrate how the theoretical developments of our work can be used to gain insights and estimate useful quantities that have potential in influencing policy. We have practiced caution and not made predictions for the future. This is because we understand that our model, and other similar models are approximations of the system and the assumptions made by the model need not be consistent over large periods of time, especially in the unforeseeable future. The parameters of the model are constantly changing as the virus continues to mutate and society continues to adapt. Thus, while such models may be powerful in understanding certain properties of the system in specific periods of time, they need not have similar utility in predicting the future. Hence, we have employed the model keeping the purpose with which it was developed in mind, that is, to understand the system better, as it is at present, and as it has been for the several months until now.

\section*{Acknowledgment}
SPC was supported by Grant No. MSC/2020/000049 from the Science and Engineering Research Board, Government of India.

\bibliographystyle{apalike}
\bibliography{Bib.bib}

\begin{thebibliography}{}

\bibitem[Ali et~al., 2020]{Ali2020}
Ali, M., Shah, S. T.~H., Imran, M., and Khan, A. (2020).
\newblock The role of asymptomatic class, quarantine and isolation in the
  transmission of covid-19.
\newblock {\em Journal of biological dynamics}, 14(1):389--408.

\bibitem[Barthelemy, 2019]{Barthelemy2019}
Barthelemy, M. (2019).
\newblock The statistical physics of cities.
\newblock {\em Nature Reviews Physics}, 1(6):406--415.

\bibitem[Blasius, 2020]{Blasius2020}
Blasius, B. (2020).
\newblock Power-law distribution in the number of confirmed covid-19 cases.
\newblock {\em Chaos: An Interdisciplinary Journal of Nonlinear Science},
  30(9):093123.

\bibitem[Bouchaud and M{\'e}zard, 2000]{Bouchaud2000}
Bouchaud, J.-P. and M{\'e}zard, M. (2000).
\newblock Wealth condensation in a simple model of economy.
\newblock {\em Physica A: Statistical Mechanics and its Applications},
  282(3-4):536--545.

\bibitem[Castilho et~al., 2020]{Castilho2020}
Castilho, C., Gondim, J.~A., Marchesin, M., and Sabeti, M. (2020).
\newblock Assessing the efficiency of different control strategies for the
  coronavirus (covid-19) epidemic.
\newblock {\em arXiv preprint arXiv:2004.03539}.

\bibitem[COVID-19~India, 2020]{DataIndia}
COVID-19~India, I. (2020).
\newblock Coronavirus outbreak in india.
\newblock {\em covid19india.org}.

\bibitem[Di~Domenico et~al., 2020]{Di2020}
Di~Domenico, L., Pullano, G., Sabbatini, C.~E., Bo{\"e}lle, P.-Y., and Colizza,
  V. (2020).
\newblock Impact of lockdown on covid-19 epidemic in {\^i}le-de-france and
  possible exit strategies.
\newblock {\em BMC medicine}, 18(1):1--13.

\bibitem[Efron and Tibshirani, 1986]{Efron1986}
Efron, B. and Tibshirani, R. (1986).
\newblock Bootstrap methods for standard errors, confidence intervals, and
  other measures of statistical accuracy.
\newblock {\em Statistical science}, pages 54--75.

\bibitem[Feng, 2007]{Feng2007}
Feng, Z. (2007).
\newblock Final and peak epidemic sizes for seir models with quarantine and
  isolation.
\newblock {\em Mathematical Biosciences \& Engineering}, 4(4):675.

\bibitem[Gibson, 2020]{Gibson2020}
Gibson, J. (2020).
\newblock Government mandated lockdowns do not reduce covid-19 deaths:
  implications for evaluating the stringent new zealand response.
\newblock {\em New Zealand Economic Papers}, pages 1--12.

\bibitem[Gupta et~al., 2021a]{Gupta2021A}
Gupta, A., Malani, A., and Woda, B. (2021a).
\newblock Inequality in india declined during covid.
\newblock Technical report, National Bureau of Economic Research.

\bibitem[Gupta et~al., 2021b]{Gupta2021B}
Gupta, A., Zhu, H., Doan, M.~K., Michuda, A., and Majumder, B. (2021b).
\newblock Economic impacts of the covid- 19 lockdown in a remittance-dependent
  region.
\newblock {\em American Journal of Agricultural Economics}, 103(2):466--485.

\bibitem[Gupta et~al., 2021c]{Gupta2021C}
Gupta, M., Mohanta, S.~S., Rao, A., Parameswaran, G.~G., Agarwal, M., Arora,
  M., Mazumder, A., Lohiya, A., Behera, P., Bansal, A., et~al. (2021c).
\newblock Transmission dynamics of the covid-19 epidemic in india and modeling
  optimal lockdown exit strategies.
\newblock {\em International Journal of Infectious Diseases}, 103:579--589.

\bibitem[He et~al., 2020]{He2020}
He, S., Peng, Y., and Sun, K. (2020).
\newblock Seir modeling of the covid-19 and its dynamics.
\newblock {\em Nonlinear dynamics}, 101(3):1667--1680.

\bibitem[Kapoor et~al., 2021]{Kapoor2021}
Kapoor, M., Ravi, S., and Kumar, A.~S. (2021).
\newblock Covid 19, consumption and inequality: a systematic analysis of rural
  population of india.
\newblock {\em medRxiv}.

\bibitem[Lakshminarayanan and Jayalakshmy, 2015]{Lakshminarayanan2015}
Lakshminarayanan, S. and Jayalakshmy, R. (2015).
\newblock Diarrheal diseases among children in india: Current scenario and
  future perspectives.
\newblock {\em Journal of natural science, biology, and medicine}, 6(1):24.

\bibitem[Ministry~of Tourism, 2020]{TIS2020}
Ministry~of Tourism, I. (2020).
\newblock India tourism statistics.
\newblock {\em
  https://tourism.gov.in/sites/default/files/2021-05/INDIA\%20TOURISM\%20STATISTICS\%202020.pdf}.

\bibitem[Ministry~of Tourism, 2021]{TIS2021}
Ministry~of Tourism, I. (2021).
\newblock India tourism statistics at a glance.
\newblock {\em
  https://tourism.gov.in/sites/default/files/2021-09/English\%20Tourisum\%202021.pdf}.

\bibitem[Pai et~al., 2017]{Pai2017}
Pai, M., Correa, N., Mistry, N., and Jha, P. (2017).
\newblock Reducing global tuberculosis deaths—time for india to step up.
\newblock {\em The Lancet}, 389(10075):1174--1176.

\bibitem[Roser et~al., 2020]{Roser2020}
Roser, M., Ritchie, H., Ortiz-Ospina, E., and Hasell, J. (2020).
\newblock Coronavirus pandemic (covid-19).
\newblock {\em Our world in data}.

\bibitem[Sharma and Mahendru, 2020]{Sharma2020}
Sharma, G.~D. and Mahendru, M. (2020).
\newblock Lives or livelihood: Insights from locked-down india due to covid19.
\newblock {\em Social Sciences \& Humanities Open}, 2(1):100036.

\bibitem[Sharma et~al., 2015]{Sharma2015}
Sharma, R.~K., Thakor, H., Saha, K., Sonal, G., Dhariwal, A., and Singh, N.
  (2015).
\newblock Malaria situation in india with special reference to tribal areas.
\newblock {\em The Indian journal of medical research}, 141(5):537.

\bibitem[UIDAI, 2020]{PopIndia}
UIDAI, U. I. A. o.~I. (2020).
\newblock State/ut wise aadhaar saturation (overall) - all age groups.
\newblock {\em https://uidai.gov.in/images/state-wise-aadhaar-saturation.pdf}.

\end{thebibliography}

\end{document}